\documentclass[journal=jacsat,manuscript=article]{achemso}
\usepackage[utf8]{inputenc}
\usepackage{color}
\usepackage{amsmath}
\usepackage{amssymb}
\usepackage{graphicx}
\usepackage{esint}
\PassOptionsToPackage{version=3}{mhchem}
\usepackage{mhchem}
\usepackage[unicode=true,pdfusetitle,
 bookmarks=true,bookmarksnumbered=false,bookmarksopen=false,
 breaklinks=false,pdfborder={0 0 1},backref=false,colorlinks=false]
 {hyperref}
\usepackage{breakurl}

\makeatletter

\providecommand{\tabularnewline}{\\}

\PassOptionsToPackage{version=3}{mhchem}

\providecommand{\tabularnewline}{\\}

\usepackage{amsthm}% Useful AMS stuff

\usepackage{tikz}\usetikzlibrary{arrows}
\usepackage{caption}\usepackage{subcaption}\usepackage{color}\usepackage{lscape}

\usepackage[section]{placeins}

\newcommand{\bra}[1]{\langle #1 \mid}
\newcommand{\ket}[1]{\mid #1 \rangle}
\newcommand{\bracket}[2]{\langle #1 \mid #2 \rangle}

\DeclareMathOperator\erf{erf}
\renewcommand{\imath}{\text{\rm{i}}}

\usepackage{titlesec}\titleformat{\chapter}
{\normalfont\huge\bfseries}{\chaptertitlename\ \thechapter:\ \ }{0em}{} 
\titlespacing*{\chapter}{0pt}{0pt}{30pt}

\newenvironment{rcases}
  {\left.\begin{aligned}}
  {\end{aligned}\right\rbrace}

\author{Alexander Humeniuk}
\author{Roland Mitri\'{c}}
\affiliation{Institut f\"{u}r Physikalische und Theoretische Chemie, Julius-Maximilians Universit\"{a}t W\"{u}rzburg, Emil-Fischer-Stra\ss e 42, 97074 W\"{u}rzburg}
\email{roland.mitric@uni-wuerzburg.de}
\title{Long-range correction for tight-binding TD-DFT}

\abbreviations{DFT, TD-DFT, DFTB, TD-DFTB}
\keywords{tight-binding DFT, long-range correction, charge transfer}

\makeatother

\begin{document}
\begin{tocentry} \includegraphics[width=1\textwidth]{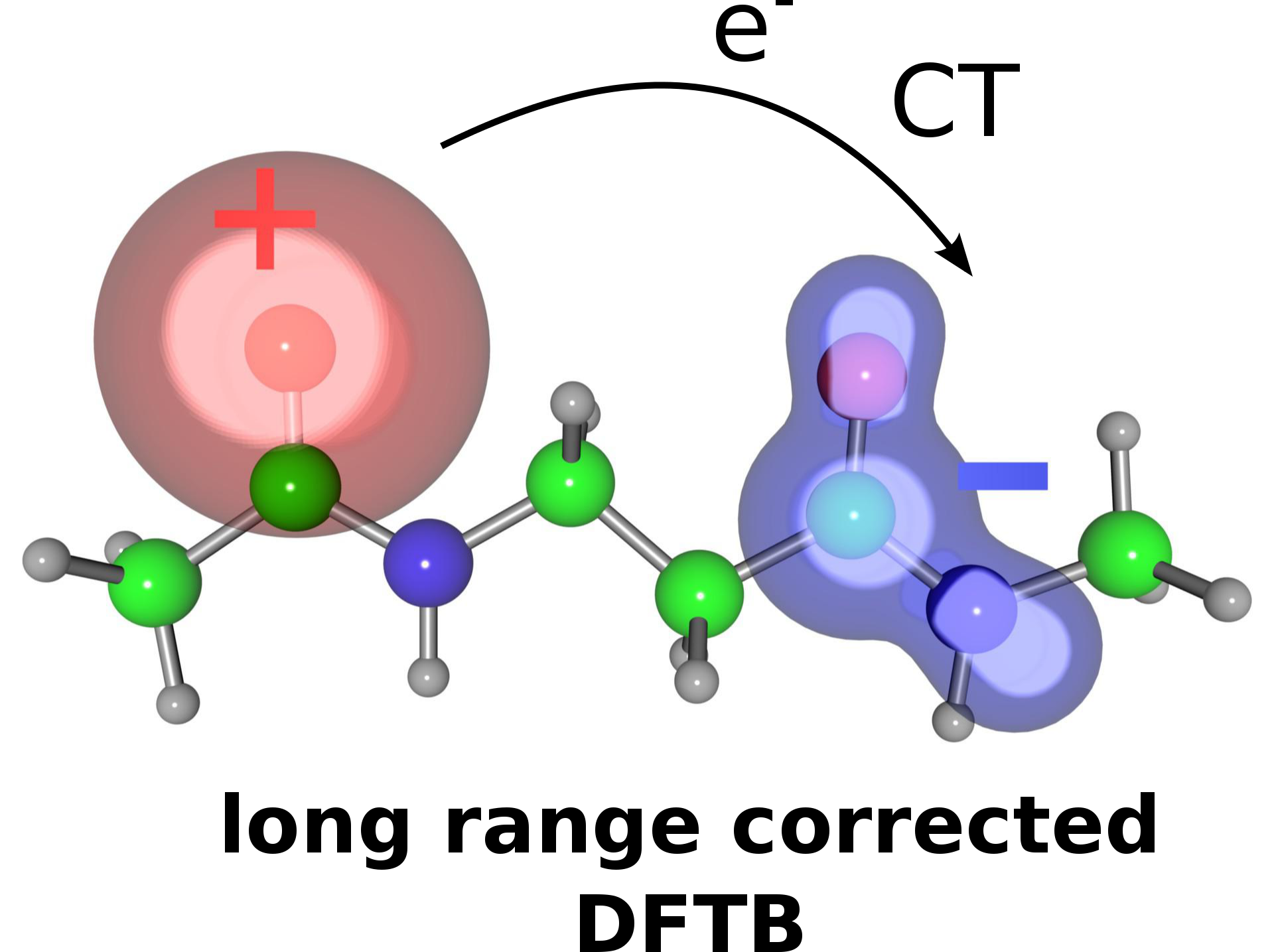} \end{tocentry}


\begin{abstract}
We present two improvements to the tight-binding approximation of
time-dependent density functional theory (TD-DFTB): Firstly, we add
an exact Hartree-Fock exchange term, which is switched on at large
distances, to the ground state Hamiltonian and similarly to the coupling
matrix that enters the linear response equations for the calculation of
excited electronic states. We show that the excitation energies of
charge transfer states are improved relative to the standard approach
without the long-range correction by testing the method on a set of
molecules from the database in {[}\textit{J. Chem. Phys.} \textbf{2008,}
\textsl{128,} 044118{]} that are known to exhibit problematic charge
transfer states. The degree of spatial overlap between occupied and
virtual orbitals indicates where TD-DFTB and lc-TD-DFTB can be expected
to produce large errors.

Secondly, we improve the calculation of oscillator strengths. The
transition dipoles are obtained from Slater Koster files for the dipole
matrix elements between valence orbitals. In particular, excitations
localized on a single atom, which appear dark when using Mulliken
transition charges, in this way acquire a more realistic oscillator
strength.

These extensions pave the way for using long-range corrected TD-DFTB
(lc-TD-DFTB) to describe the electronic structure of large chromophoric
polymers, where uncorrected TD-DFTB fails to describe the high degree
of conjugation and produces spurious low-lying charge transfer states. 
\end{abstract}
\maketitle

\section{Introduction}

Tight-binding DFT (DFTB)\cite{DFTB_elstner,DFTB_large_biomolecules,DFTB_for_beginners}
and its time-dependent formulation TD-DFTB\cite{TD-DFTB_Niehaus_Elstner_Fraunheim,TD-DFTB_Barone}
are semi-empirical methods based on (TD)-DFT\cite{TDDFRT_Casida,TDDFT_Casida,TDDFT_Furche}
that inherit many of the advantages and shortcomings of the latter.
The failure of TD-DFT to describe charge transfer states\cite{TDDFT_spurious_CT}
is particularly severe if one deals with extended molecules or oligomers
with a large degree of conjugation. These charge transfer states,
which appear at unphysically low energies, can be removed if a long-range
exchange term is included, leading to the long-range corrected
TD-DFT \cite{long_range_corrected_TDDFT}.

In the tight-binding formulation this correction can also be included
at almost no additional computational cost. The possibility to include
a range separated functional into DFTB has been explored before by
Niehaus and Della Sala \cite{range_separated_DFTB}. They generate
the pseudo atom basis starting from a full DFT calculation on a single
atom with a range-separated functional. This changes the electronic
parametrization and makes the pseudoatoms depend on the long-range
correction. %repulsive potentials be readjusted and validated again, which is the most time-consuming part in the parametrization of DFTB. 
Then they consistently add the first order correction arising from
the long-range correction in the tight-binding Kohn-Sham equations
and the linear response formalism.

Here instead, a much simpler (maybe less rigorous) approach is proposed,
where the existing electronic parameters (pseudo atoms and Hubbard
parameters) are left untouched. Since the usual DFTB parametrizations
are based on a local density approximation (LDA), the long range exchange
can be incorporated by simply adding the attenuated exact exchange
energy $E_{x}^{\text{lr}}$ to the total electronic ground state energy.
For the calculation of excited states, a long range correction term
is added to the coupling matrix in the spirit of CAM-B3LYP\cite{CAM_B3LYP}.
The exchange integrals are then approximated by products of transition
charges as usual. The distance at which the exact exchange is gradually
switched on is controlled by new a parameter $R_{\text{lr}}$ (equivalent
to $1/\mu$ in the notation of ref. \cite{long_range_corrected_TDDFT})
that can be adjusted to fit excitation energies to CAM-B3LYP results
or experimental values.

This article is structured as follows: In sections \ref{sec:lc_tddftb}
and \ref{sec:sk_tdip} the working formulae for the long-range correction
and the transition dipoles are derived. Section \ref{sec:lambda_diagnostics}
deals with different ways of quantifying charge transfer at the level
of tight-binding DFT. Finally the method is applied to a test set
of molecules and compared to CAM-B3LYP results in sections \ref{sec:computational_details}
and \ref{sec:results}.

Technical details are deferred to the appendices: 
\begin{itemize}
\item Appendix \ref{sec:appendix_gamma_matrix} elucidates ambiguities in the definition of the $\gamma$-matrix that lead to differences between DFTB implementations.
\item Appendix \ref{sec:appendix_switching_functions} discusses different ways of splitting the Coulomb potential into a short and a long range part.
\item Slater-Koster rules for transition dipoles are derived in the appendices
\ref{sec:dipoles_technical_details},
 \ref{sec:unique_radial_integrals} and
 \ref{sec:slako_rules_dipoles}.

% \ref{sec:ct_analysis_drho}
\end{itemize}

\section{Method Description}

Many review articles have been written about DFTB\cite{DFTB_review_brazil,DFTB_review_Elstner_Seifert},
for a pedagogical introduction see \cite{DFTB_for_beginners}. To
make clear at which points we introduce modifications, we recapitulate here
the basics of DFTB and TD-DFTB:

%%%%%%%%%%%%%%%%%%%%%%%%%%%%%%%%%%%%%%%%%%%%%%%%%%
To derive the DFTB Hamiltonian, one starts with the full DFT energy
functional\cite{kohn_sham} 
\begin{equation}
E[\rho]=\sum_{\alpha}f_{a}\bra{\phi_{\alpha}}\left(-\frac{1}{2}\nabla^{2}+\int V_{\text{ext}} \rho \right)\ket{\phi_{\alpha}}+\frac{1}{2}\int\int'\frac{\rho\rho'}{\vert\vec{r}-\vec{r}'\vert}+E_{\text{xc}}[\rho]+E_{\text{nuc}}
\end{equation}
where $\phi_{\alpha}$ are the Kohn-Sham orbitals of the system of
non-interacting electrons with the occupation numbers $f_{\alpha}$,
$V_{\text{ext}}$ is the potential the nuclei exert on the electrons,
as well as any external electric field, $E_{xc}$ is the exchange-correlation
functional and $E_{\text{nuc}}$ the energy due to Coulomb
repulsion between nuclei. % V_H - Hartree potential, V_xc - exchange correlation potential, 
One proceeds by expanding the energy around a reference density, $\rho=\rho_{0}+\delta\rho$,
to second order in $\delta\rho$. The reference density $\rho_{0}$
is a superposition of atomic densities of the individually neutral
atoms, while the redistribution of the electron density $\delta\rho$
is a result of the chemical bonding.

After rearranging the expansion of the energy functional, 
\begin{equation}
\begin{split}E[\rho_{0}+\delta\rho]\approx & \sum_{\alpha}f_{\alpha}\bra{\phi_{\alpha}}\left(-\frac{1}{2}\nabla^{2}+V_{\text{ext}}+V_{H}[\rho_{0}]+V_{\text{xc}}[\rho_{0}]\right)\ket{\phi_{\alpha}}\\
 & +\frac{1}{2}\int\int'\left(\frac{\delta^{2}E_{xc}[\rho_{0}]}{\delta\rho\delta\rho'}+\frac{1}{\vert\vec{r}-\vec{r}'\vert}\right)\delta\rho\delta\rho'\\
 & -\frac{1}{2}\int V_{H}[\rho_{0}]\rho_{0}(\vec{r})+\left(E_{xc}[\rho_{0}]-\int V_{xc}[\rho_{0}]\rho_{0}(\vec{r})\right)+E_{\text{nuc}},
\end{split}
\end{equation}
the energy is partitioned into terms depending on the reference density
and the orbitals ($E_{\text{bs}}$), the density fluctuations ($E_{\text{coul,xc}}$)
and the repulsive energy ($E_{\text{rep}}$) , which stands for everything else not covered
by the first two terms: 
\begin{equation}
\begin{split}E[\rho_{0}+\delta\rho]\approx & \sum_{a}f_{a}\bra{\phi_{a}}H[\rho_{0}]\ket{\phi_{a}}\quad\quad\quad\quad\quad\quad\quad\quad\quad\quad\quad\text{band structure energy }E_{bs}\\
 & +\frac{1}{2}\int\int'\left(\frac{\delta^{2}E_{xc}[\rho_{0}]}{\delta\rho\delta\rho'}+\frac{1}{\vert\vec{r}-\vec{r}'\vert}\right)\delta\rho\delta\rho'\quad\quad\quad\text{Coulomb + part of xc energy }E_{\text{coul,xc}}\\
 & +E_{\text{nuc}}+\text{ everything else}\quad\quad\quad\quad\quad\quad\quad\quad\quad\text{repulsive energy }E_{\text{rep}}
\end{split}
\end{equation}

To approximate $E_{\text{coul,xc}}$ one assumes that the charge fluctuation
$\delta\rho$ can be decomposed into spherically symmetric contributions
centered on the atoms, $\delta\rho=\sum_{I}^{N_{at}}\Delta q_{I}F_{I}(\vert\vec{r}-\vec{R}_{I}\vert)$,
where $\Delta q_{I}$ are the excess Mulliken charges on atom $I$:
\begin{equation}
\begin{split}E_{\text{coul,xc}}[\rho_{0}+\delta\rho] & =\frac{1}{2}\int\int'\left(\frac{\delta^{2}E_{xc}[\rho_{0}]}{\delta\rho\delta\rho'}+\frac{1}{\vert\vec{r}-\vec{r}'\vert}\right)\delta\rho\delta\rho'\\
 & =\sum_{I}^{N_{at}}\sum_{J}^{N_{at}}E_{\text{coul,xc}}^{IJ}\label{eqn:Ecoulxc}
\end{split}
\end{equation}
For partial charges sitting on different atoms, $I\neq J$, only the
electrostatic interaction is taken into account (depending on the
distance $R_{IJ}$ of the atomic centers) and any exchange or correlation
interaction is neglected, since the exchange correlation functional
is assumed to be local. The interaction energy of charge with itself
on the same atom is controlled by the Hubbard parameters $U_{H}$,
which can be obtained from experimental ionization energies and electron
affinities, or ab-initio calculations.
This leads to a partitioning of $E_{\text{coul,xc}}^{IJ}$ into pievewise contributions:
\begin{equation}
E_{\text{coul,xc}}^{IJ}=\left.\begin{cases}
\frac{1}{2}\Delta q_{I}\Delta q_{J}\int\int'\frac{F_{I}F_{J}'}{\vert\vec{r}-\vec{r}'\vert} & I\neq J\\
\frac{1}{2}U_{H}\left(\Delta q_{I}\right)^{2} & I=J
\end{cases}\right\} =\frac{1}{2}\Delta q_{I}\Delta q_{J}\gamma_{IJ}(R_{IJ})\label{eqn:Ecoulxc_electrostatic}
\end{equation}

The approximate DFTB energy is now a function of the partial Mulliken
charges instead of the density: 
\begin{equation}
E_{\text{DFTB}}[\{\Delta q_{I}\}]=\sum_{\alpha}f_{\alpha}\bra{\phi_{\alpha}}H[\rho_{0}]\ket{\phi_{a}}+\frac{1}{2}\sum_{IJ}\gamma_{IJ}(R_{IJ})\Delta q_{I}\Delta q_{J}+\underbrace{\left(E[\rho]-E_{\text{DFTB}}\right)}_{\sum_{I<J}V_{\text{rep}}^{IJ}(R_{IJ})}
\end{equation}
All the deviations from the true energy are absorbed into the \textsl{repulsive
potential}, which ideally should only depend on the molecular geometry
but not on the charge distribution. In a rough approximation, these
deviations can be decomposed into contributions from pairs of atoms, $V_{\text{rep}}^{IJ}(R_{IJ})$,
and be adjusted to higher-level DFT methods.

Fitting\cite{parametrization_dftb3} and validating\cite{dftb_validation}
the repulsive potentials is the most time-consuming part of parametrizing
the DFTB method. Although properly adjusted repulsive potentials are
crucial for molecular dynamics simulations, structure optimization
or vibrational spectra, we will neglect them here, as they have no
influence on the electronic absorption spectra.

The Kohn-Sham orbitals are expanded into a minimal set of pseudo-atomic
orbitals $\{\ket{\mu}\}$, which are compressed by a confining potential,
since orbitals in a molecule are less diffuse than in the free atoms:
\begin{equation}
\ket{\phi_{i}}=\sum_{\mu}c_{\mu}^{i}\ket{\mu}
\end{equation}

A variation of the energy with respect to the Kohn-Sham orbitals, under
the constraint that the orbitals are normalized, leads to the DFTB
equivalent of the Kohn-Sham equations with the DFTB Hamiltonian (in
the basis of atomic orbitals $\mu$ and $\nu$): 
\begin{equation}
H_{\mu\nu}^{\text{DFTB}}=\underbrace{\bra{\mu}H[\rho_{0}]\ket{\nu}}_{H_{\mu\nu}^{0}}+\frac{1}{2}\underbrace{\bracket{\mu}{\nu}}_{S_{\mu\nu}}\sum_{K=1}^{N_{at}}\left(\gamma_{IK}+\gamma_{JK}\right)\Delta q_{K}\quad\quad\mu\in I,\nu\in J
\end{equation}
where $\mu\in I$ means that the atomic orbital $\mu$ belongs to
atom $I$.

% charge-constistency loop 
The Kohn-Sham equations are solved self-consistently: In each step
the new partial Mulliken charges $\Delta q_{I}$ and the new Hamiltonian
are computed from the orbital coefficients of the previous iteration
and the resulting Kohn-Sham equations are solved to give the next
orbital coefficients. These steps are repeated until the charge distribution
and the density matrix do not change anymore.

Excited states are calculated in the framework of linear-response
TD-DFT, which was adapted to tight-binding DFT by Niehaus\cite{TD-DFTB_Niehaus_Elstner_Fraunheim}.
For brevity, we only sketch the working equations of TD-DFT and how
they are adapted to the tight-binding formulation. A converged DFT
calculation provides the single-particle Kohn-Sham orbitals, and to
a first approximation excitation energies are differences of virtual
and occupied Kohn-Sham energies $\omega_{ov\sigma}=\epsilon_{v\sigma}-\epsilon_{o\sigma}$.

The coupling matrix 
\begin{equation}
K_{ov\sigma,o'v'\sigma'}=\int\int\phi_{o \sigma}(\vec{r}_{1})\phi_{v \sigma}(\vec{r}_{1})\left[\frac{1}{\vert\vec{r}_{1}-\vec{r}_{2}\vert}+\frac{\delta^{2}E_{xc}}{\delta\rho_{\sigma}(\vec{r}_{1})\delta\rho_{\sigma'}(\vec{r}_{2})}\right]\phi_{o' \sigma'}(\vec{r}_{2})\phi_{v' \sigma'}(\vec{r}_{2})d^{3}r_{1}d^{3}r_{2}
\end{equation}
represents the response of the Kohn-Sham potential to a perturbation
of the electron density and is responsible for adding a correction
to the Kohn-Sham orbital energy differences.

%############################

In linear-response TD-DFT excited states are computed from the Hermitian
eigenvalue problem\cite{TDDFT_Casida}
\begin{equation}
(\mathbf{A}-\mathbf{B})^{1/2}(\mathbf{A}+\mathbf{B})(\mathbf{A}-\mathbf{B})^{1/2}\vec{F}_{I}=\Omega_{I}^{2}\vec{F}_{I}\label{eqn:hermitian_eigval}
\end{equation}
where the matrices $\mathbf{A}$ and $\mathbf{B}$ contain the coupling
matrix: 
\begin{eqnarray}
A_{ov\sigma,o'v'\sigma'} & = & \delta_{o,o'}\delta_{v,v'} \delta_{\sigma,\sigma'} (\epsilon_{v\sigma}-\epsilon_{o\sigma})+K_{ov\sigma,o'v'\sigma'}\\
B_{ov\sigma,o'v'\sigma'} & = & K_{ov\sigma,v'o'\sigma'}
\end{eqnarray}
The solution of eqn. (\ref{eqn:hermitian_eigval}) provides the excitation
energies $E_{I}=\hbar\Omega_{I}$ as eigenvalues and the coefficients
for single excitations from occupied to virtual orbitals $(o\to v)$
\begin{equation}
\vec{C}_{ov\sigma}^{I}=\frac{1}{\sqrt{\Omega_{I}}}\sum_{o'\in occ}\sum_{v'\in virt} \sum_{\sigma'} \left[(\mathbf{A}-\mathbf{B})^{1/2}\right]_{ov\sigma,o'v'\sigma'}\vec{F}_{o'v'\sigma'}^{I}
\end{equation}
as eigenvectors.

In the absence of a non-local exchange term, $(\mathbf{A}-\mathbf{B})$
is effectively diagonal so that eqn. (\ref{eqn:hermitian_eigval})
can be simplified to yield Casida's equation\cite{TDDFT_Casida}:

\begin{equation}
\sum_{o'\in occ}\sum_{v'\in virt} \sum_{\sigma'} \left[\delta_{o,o'}\delta_{v,v'} \delta_{\sigma,\sigma'} \left(\epsilon_{v\sigma}-\epsilon_{o\sigma}\right)^{2}+2\sqrt{\epsilon_{v\sigma}-\epsilon_{o\sigma}}K_{ov\sigma,o'v'\sigma'}\sqrt{\epsilon_{v'\sigma'}-\epsilon_{o'\sigma'}}\right]F_{o'v'\sigma'}^{I}=\Omega_{I}^{2}F_{ov\sigma}^{I} \label{eqn:Casidas_equation}
\end{equation}
with the coefficients 
\begin{equation}
C_{ov\sigma}^{I}=\sqrt{\frac{\epsilon_{v\sigma}-\epsilon_{o\sigma}}{\Omega_{I}}}F_{ov\sigma}^{I}.
\end{equation}

In the Tamm-Dancoff approximation\cite{TDDFT_Tamm_Dancoff,TDDFT_Tamm_Dancoff_Martinez}
excitation energies and coefficients are calculated from a different
eigenvalue problem that results from eqn. (\ref{eqn:hermitian_eigval})
by setting $\mathbf{B}=0$: 
\begin{equation}
\sum_{o'\in occ}\sum_{v'\in virt} \sum_{\sigma'} A_{ov\sigma,o'v'\sigma'}C_{o'v'\sigma'}^{I}=\Omega_{I}C_{ov\sigma}^{I}
\end{equation}

Using the Tamm-Dancoff (TDA) approximation 
in conjunction with a long-range correction (which will be introduced
later) can sometimes be advantageous for two reasons: 
\begin{itemize}
\item Firstly, it has been shown \cite{Casida_TDA_oxirane,TDDFT_singlet_triplet}
that TDA excitation energies can actually be better than those obtained
from the full solution of the LR-TD-DFT equation \ref{eqn:hermitian_eigval},
in particular, when singlet-triplet instabilities would lead to imaginary
excitation energies. 
\item Secondly, when $(\mathbf{A}-\mathbf{B})$ is not diagonal, the full
solution requires the computation of the matrix square root $(\mathbf{A}-\mathbf{B})^{1/2}$
which is computationally demanding unless one resorts to an iterative
algorithm\cite{TDDFT_efficient} specifically designed to deal with
eqn. (\ref{eqn:hermitian_eigval}). 
\end{itemize}
%############################

The tight-binding approximation consists in replacing transition densities
$\phi_{o}(\vec{r})\phi_{v}(\vec{r})$ by transition charges $q_{A}^{ov}$
(defined later), so that the coupling matrix simplifies to 
\begin{equation}
K_{ov,o'v'}\approx\sum_{A=1}^{N_{at}}\sum_{B=1}^{N_{at}}q_{A}^{ov}\gamma_{AB}q_{B}^{o'v'}.\label{eqn:DFTB_coupling_matrix}
\end{equation}
Similarly, the transition dipoles between Kohn-Sham orbitals are reduced
to sums over transition charges on different atoms: 
\begin{equation}
\bra{o}\vec{r}\ket{v}\approx\sum_{A}\vec{R}_{A}q_{A}^{ov}\label{eqn:oscillator_strengths_DFTB}
\end{equation}

Note how in going from eqn. (\ref{eqn:Ecoulxc}) to the eqn. (\ref{eqn:Ecoulxc_electrostatic})
and in approximating the coupling matrix in eqn. (\ref{eqn:DFTB_coupling_matrix})
the exchange-correlation term has been neglected for charge distributions
on different atoms, arguing that the xc-functional is local. This
is where the long-range correction will be put to work. Eqn. (\ref{eqn:oscillator_strengths_DFTB})
is usually a reasonable approximation, unless orbitals $o$ and $v$
are located on the same atom.

%%%%%%%%%%%%%%%%%%%%%%%%%%%%%%%%%%%%%%%%%%%%%%%%%%%%%%

\subsection{Long-range correction for TD-DFTB}

\label{sec:lc_tddftb} The Coulomb potential is separated into a long-range
and a short-range part\cite{LC_Savin, LC_Gaussian}, where the position of the smooth transition
between the two regimes is controlled by a parameter $R_{\text{lr}}$:
\begin{equation}
\frac{1}{r}=\underbrace{\frac{1-\erf\left(\frac{r}{R_{\text{lr}}}\right)}{r}}_{\text{short range}}+\underbrace{\frac{\erf\left(\frac{r}{R_{\text{lr}}}\right)}{r}}_{\text{long range}}
\end{equation}
The short range part of the exchange energy is treated with DFTB while
for the long range part the exact Hartree-Fock exchange is used. Since
in DFTB a local exchange correlation functional is employed, the short
range term is essentially neglected.

The electron integrals of the screened Coulomb potential (for real-valued orbitals) 
\begin{equation}
(ij\vert ab)_{\text{lr}}=\int\int\phi_{i}(\vec{r}_{1})\phi_{j}(\vec{r}_{1})\frac{\erf\left(\frac{r_{12}}{R_{\text{lr}}} \right)}{r_{12}}\phi_{a}(\vec{r}_{2})\phi_{b}(\vec{r}_{2})d^{3}r_{1}d^{3}r_{2}
\end{equation}
are approximated as in DFTB \cite{DFTB_for_beginners}: The transition
densities between different orbitals $p^{kl}(\vec{r})=\phi_{k}(\vec{r})\phi_{l}(\vec{r})$
are decomposed into atom-centered contributions: 
\begin{equation}
p^{kl}(\vec{r})=\sum_{A}^{N_{\text{at}}}p_{A}^{kl}(\vec{r})
\end{equation}
Next, the monopole approximation is made assuming that the transition
density due to atom $A$ is spherically symmetric around that center:
\begin{equation}
p_{A}^{kl}(\vec{r})=q_{A}^{kl}F_{A}(\vert\vec{r}-\vec{R}_{A}\vert)
\end{equation}

In fact, the exact form of the functions $F_{A}(r)$ is not know\cite{DFTB_for_beginners}. They can be assumed to be Gaussian\cite{DFTB_for_beginners} or Slater functions\cite{range_separated_DFTB}, but in either case the width of the density profile should be inversely proportional to the chemical hardness (the Hubbard parameter $U$) of the atom.

With these approximations the long-range two-center integrals can be written
in terms of transition charges $q_{A}^{ij}$ and $q_{B}^{ab} $: 
\begin{equation}
(ij\vert ab)_{\text{lr}}=\sum_{A}\sum_{B}q_{A}^{ij}q_{B}^{ab}\int\int
F_{A}(\vert\vec{r}_{1}-\vec{R}_{A}\vert)
\frac{\erf\left(\frac{r_{12}}{R_{\text{lr}}} \right)}{r_{12}}
F_{B}(\vert\vec{r}_{2}-\vec{R}_{B}\vert)
d^{3}r_{1}d^{3}r_{2} \label{eqn:gamma_lr_integral}
\end{equation}
The transition charges are defined as: 
\begin{equation}
q_{A}^{ij}=\frac{1}{2}\sum_{\mu\in A}\sum_{\nu}\left(c_{\mu}^{i}c_{\nu}^{j}+c_{\nu}^{i}c_{\mu}^{j}\right)S_{\mu\nu}.
\end{equation}
where $S_{\mu\nu}$ is the overlap matrix between the atomic orbitals
$\mu$ and $\nu$ and $c_{\mu}^{i}$ is the coefficient of the atomic
orbital $\mu$ in the molecular orbital $i$.

Assuming a Gaussian- or a Slater-function form for $F_{A}(r)$ the integral for the unscreened Coulomb potential can be performed
analytically \cite{DFTB_for_beginners} and the result
is defined as the $\gamma$-matrix: 
\begin{equation}
\gamma_{AB}=\int\int\frac{F_{A}(\vert\vec{r}_{1}-\vec{R}_{A}\vert)F_{B}(\vert\vec{r}_{2}-\vec{R}_{B}\vert)}{\vert\vec{r}_{1}-\vec{r}_{2}\vert}d^{3}r_{1}d^{3}r_{2} \label{eqn:gamma_Coulomb_integral}
\end{equation}

For the long range part of the Coulomb potential the error-function
can be taken out of the integral in eqn. \ref{eqn:gamma_lr_integral} 
in a good approximation, giving the long-range $\gamma$-matrix 
\begin{equation}
\gamma_{AB}^{\text{lr}}(R_{AB}) \approx \erf(\frac{R_{AB}}{R_{\text{lr}}})\gamma_{AB}(R_{AB}) \label{eqn:gamma_lr_approx}
\end{equation}
from which the electron integrals with the long range part of the
Coulomb potential can be calculated as: 
\begin{equation}
(\mu\lambda\vert\sigma\nu)_{\text{lr}}\approx\sum_{A}\sum_{B}\gamma_{AB}^{\text{lr}}q_{A}^{\mu\lambda}q_{B}^{\sigma\nu}
\end{equation}

\begin{figure}[h!]
\includegraphics[width=0.8\textwidth]{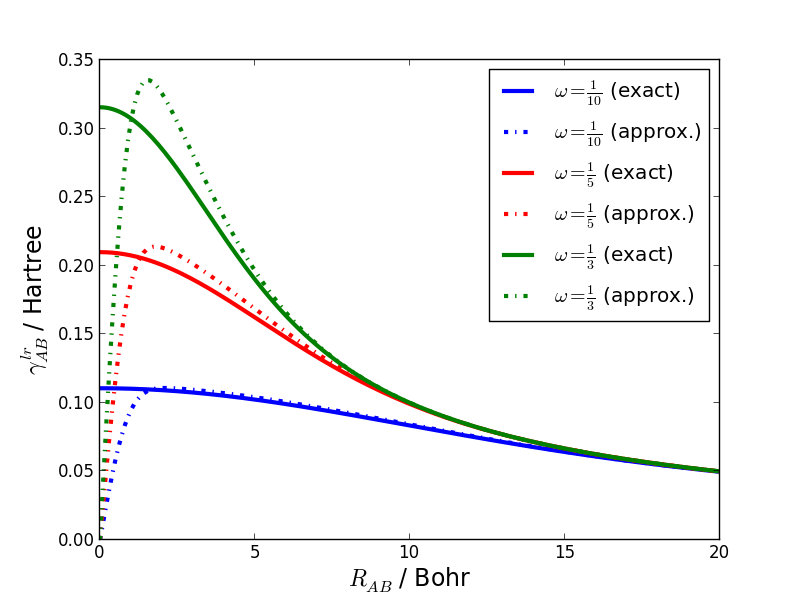}
\caption{Comparison between the exact solution of the integral in eqn. \ref{eqn:gamma_lr_integral} and the approximation made in eqn. \ref{eqn:gamma_lr_approx} for different $\omega = \frac{1}{R_{lr}}$. The green curves are plotted for the long-range radius $R_{lr} = 3.0$ bohr that is actually used in our parametrization. The approximation is quite accurate for interatomic distances that are relevant for charge transfer. For short atomic distances our approximation slightly overestimates $\gamma^{lr}_{AB}$ and vanishes for $R_{AB} \to 0$. 
In particular the diagonal elements of the Hamiltonian $H_{\mu\nu}$ are not modified, since $R_{AB} = 0$ if the orbitals $\mu$ and $\nu$ are located on the same atom.
}
\label{fig:comparison_gamma_lr_exact_approx} 
\end{figure}

%%%%%%%%%%%%%%
Incidentally, the authors of \cite{range_separated_DFTB} have shown that the integral \ref{eqn:gamma_lr_integral} can be transformed into a one-dimensional integral that can be solved numerically (assuming the $F_A(r)$'s are Slater functions). As can be seen in Fig. \ref{fig:comparison_gamma_lr_exact_approx} the approximate $\gamma_{AB}^{\text{lr}}$ and the numerically exact solution differ in the limit $R_{AB} \to 0$ but coincide for larger distances. 

%%%%%%%%%%%%%%

For atomic orbitals the transition densities are simply: 
\begin{equation}
q_{A}^{\mu\lambda}=\frac{1}{2}\left(\delta(\mu\in A)+\delta(\lambda\in A)\right)S_{\mu\lambda}
\end{equation}
Here $\delta(\mu\in A)$ is equal to 1 if the atomic orbital $\mu$
is centered on the atom $A$ and 0 otherwise.

The long-range electron integrals in the basis of atomic orbitals
can now be approximated as: 
\begin{equation}
\begin{split}(\mu\lambda\vert\sigma\nu)_{\text{lr}} & \approx\sum_{A}\sum_{B}\gamma_{AB}^{\text{lr}}q_{A}^{\mu\lambda}q_{B}^{\sigma\nu}=S_{\mu\lambda}S_{\nu\sigma}\sum_{A}\sum_{B}\frac{1}{4}\gamma_{AB}^{\text{lr}}\left(\delta(\mu\in A)+\delta(\lambda\in A)\right)\left(\delta(\sigma\in B)+\delta(\nu\in B)\right)\\
 & =\frac{1}{4}S_{\mu\lambda}S_{\nu\sigma}\sum_{A}\sum_{B}\gamma_{AB}^{\text{lr}}\left\{ \delta(\mu\in A)\delta(\sigma\in B)+\delta(\mu\in A)\delta(\nu\in B)\right.\\
 & \quad\quad\quad\quad\quad\quad\quad\quad\quad\quad\left.+\delta(\lambda\in A)\delta(\sigma\in B)+\delta(\lambda\in A)\delta(\nu\in B)\right\} \\
 & =\frac{1}{4}S_{\mu\lambda}S_{\nu\sigma}\left\{ \Gamma_{\mu\sigma}^{\text{lr}}+\Gamma_{\mu\nu}^{\text{lr}}+\Gamma_{\lambda\sigma}^{\text{lr}}+\Gamma_{\lambda\nu}^{\text{lr}}\right\} \label{eqn:electron_integrals_approx}
\end{split}
\end{equation}
where the abbreviation $\Gamma_{\mu\sigma}^{\text{lr}}=\sum_{A}\sum_{B}\gamma_{AB}^{\text{lr}}\delta(\mu\in A)\delta(\sigma\in B)$
has been introduced.

The additional contribution to the total energy $E_{\text{elec}}=E_{\text{DFTB}}+E_{x}^{\text{lr}}$
due to the long-range exchange is: 
\begin{equation}
E_{x}^{\text{lr}} = - \frac{1}{4} \sum_{\mu,\lambda,\sigma,\nu} P_{\mu \sigma} P_{\lambda \nu} (\mu \lambda \vert \sigma \nu)_{\text{lr}} \label{eqn:Exlr}
\end{equation}
where $\mu,\lambda,\sigma$ and $\nu$ enumerate atomic orbitals, and the density matrix elements $P_{\mu \sigma}$ are computed from the molecular orbital coefficients $C_{\mu,i}$ as $P_{\mu,\sigma} = 2 \sum_{i \in \text{occ}} C_{\mu i} C_{\sigma i}^*$.

Minimization of the total energy with respect to the density matrix leads to the Kohn-Sham Hamiltonian
\begin{equation}
H_{\mu\nu}^{\text{KS}}=H_{\mu\nu}^{\text{DFTB}}+H_{\mu\nu}^{x,\text{lr}}
\end{equation}
with an additional term:
\begin{equation}
\begin{split}H_{\mu\nu}^{x,\text{lr}} & =-\frac{1}{2}\sum_{\lambda\sigma}P_{\lambda\sigma}(\mu\lambda\vert\sigma\nu)_{\text{lr}}\\
 & \stackrel{(\ref{eqn:electron_integrals_approx})}{=} -\frac{1}{8}\sum_{\lambda\sigma}P_{\lambda\sigma}S_{\mu\lambda}S_{\nu\sigma}\left\{ \Gamma_{\mu\sigma}^{\text{lr}}+\Gamma_{\mu\nu}^{\text{lr}}+\Gamma_{\lambda\sigma}^{\text{lr}}+\Gamma_{\lambda\nu}^{\text{lr}}\right\} \\
 & =-\frac{1}{8}\left\{ \sum_{\sigma}\left(\Gamma_{\mu\sigma}^{\text{lr}}\left(\sum_{\lambda}S_{\mu\lambda}P_{\lambda\sigma}\right)\right)S_{\sigma\nu}\right.\\
 & \quad\quad\quad+\Gamma_{\mu\nu}^{\text{lr}}\sum_{\sigma}\sum_{\lambda}\left(S_{\mu\lambda}P_{\lambda\sigma}\right)S_{\sigma\nu}\\
 & \quad\quad\quad+\sum_{\sigma}\sum_{\lambda}\left(S_{\mu\lambda}\left(P_{\lambda\sigma}\Gamma_{\lambda\sigma}^{\text{lr}}\right)\right)S_{\sigma\nu}\\
 & \left.\quad\quad\quad+\sum_{\lambda}S_{\mu\lambda}\left(\left(\sum_{\sigma}P_{\lambda\sigma}S_{\sigma\nu}\right)\Gamma_{\lambda\nu}^{\text{lr}}\right)\right\} 
\end{split}
\end{equation}
It is important to perform the sums over the indices in such an order
that they can be implemented efficiently by nested matrix multiplications.

The long-range contribution to the exchange energy in eqn. \ref{eqn:Exlr} can be computed as:

\begin{equation}
E_{x}^{\text{lr}} = - \frac{1}{8} \left\{
 \sum_{\mu,\sigma} \left( \sum_{\lambda} S_{\mu \lambda} \left( \sum_{\nu} P_{\lambda \nu} S_{\nu \sigma} \right) \right) P_{\mu\sigma} \Gamma^{\text{lr}}_{\mu\sigma} 
+\sum_{\mu,\sigma} \left(\sum_{\lambda} S_{\mu \lambda} P_{\lambda \sigma} \right) \left( \sum_{\nu} P_{\mu \nu} S_{\nu \sigma} \right)  \Gamma^{\text{lr}}_{\mu\sigma} 
 \right\}
\end{equation}

%#####################################
In the linear response formulation of TD-DFT the long-range correction
leads to an additional term in the coupling matrix, which shifts the
excitation energies of the charge transfer states up. The corrections
to the $\mathbf{A}$- and $\mathbf{B}$-matrices read (after separating the problem into separate singlet and triplet cases \cite{Joergensen_TDHF}):

%%%%%%%%%%%%%%%%%%%
\begin{equation}
\begin{rcases}
^SA_{ov,o'v'} & =  \delta_{o,o'}\delta_{v,v'}(\epsilon_{v}-\epsilon_{o})+2 K_{ov,o'v'}+K_{ov,o'v'}^{\text{lr}}\\
^SB_{ov,o'v'} & =  2 K_{ov,v'o'}+K_{ov,v'o'}^{\text{lr}}
\end{rcases} \text{for singlets}
\end{equation}
and
\begin{equation}
\begin{rcases}
^TA_{ov,o'v'} & =  \delta_{o,o'}\delta_{v,v'}(\epsilon_{v}-\epsilon_{o}) + K_{ov,o'v'}^{\text{lr}}\\
^TB_{ov,o'v'} & =  K_{ov,v'o'}^{\text{lr}}
\end{rcases} \text{for triplets},
\end{equation}

%\begin{eqnarray}
%A_{ov,o'v'} & = & \delta_{o,o'}\delta_{v,v'}(\epsilon_{v}-\epsilon_{o})+K_{ov,o'v'}+K_{ov,o'v'}^{\text{lr}}\\
%B_{ov,o'v'} & = & K_{ov,v'o'}+K_{ov,v'o'}^{\text{lr}}
%\end{eqnarray}

where the additional long-range coupling is given by: 
\begin{eqnarray}
K_{ov,o'v'}^{\text{lr}} & =-(oo'\vert vv')_{\text{lr}}\approx-\sum_{A}\sum_{B}q_{A}^{oo'}\gamma_{AB}^{\text{lr}}\left(R_{AB}\right)q_{B}^{vv'}\\
K_{ov,v'o'}^{\text{lr}} & =-(ov'\vert vo')_{\text{lr}}\approx-\sum_{A}\sum_{B}q_{A}^{ov'}\gamma_{AB}^{\text{lr}}\left(R_{AB}\right)q_{B}^{o'v}.
\end{eqnarray}

$o,o'$ are occupied and $v,v'$ are unoccupied molecular orbitals.
In addition to the transition charges between occupied and unoccupied
orbitals, $q_{A}^{ov}$, which are needed for constructing the coupling
matrix anyway, one has to calculate transition charges between occupied-occupied
orbitals, $q_{A}^{oo'}$, and between virtual-virtual orbitals $q_{B}^{vv'}$.

% Comment about triplet states
In this approximation the quality of triplet excitation energies is expected to be much lower, 
as they are essentially equal to differences between Kohn-Sham orbital energies. 
Triplet states will be left aside in this work, since spin-unrestricted DFTB\cite{tddftb_extensions} 
is necessary to describe them quantitatively and since they are dark in absorption spectra.

% Oscillator strengths
The oscillator strengths of singlet states are obtained as 
\begin{equation}
f^{I}=\frac{4}{3}\left\vert \sum_{o\in occ}\sum_{v\in virt}\bra{o}\vec{r}\ket{v}\sqrt{\Omega_{I}}C_{ov}^{I}\right\vert ^{2}.
\end{equation}

\subsection{Slater-Koster rules for dipole matrix elements}

\label{sec:sk_tdip}

%\subsubsection{When the Mulliken approximation fails}

Transition dipole matrix elements determine the oscillator strengths
in TD-DFT calculations of excited states. In DFTB they are usually
approximated by transition charges located on the individual atoms,
\begin{equation}
\bra{\psi^{i}}\vec{r}\ket{\psi^{j}}=\sum_{\alpha}\vec{R}_{\alpha}q_{\alpha}^{ij},
\end{equation}
where $\vec{R}_{\alpha}$ is the position vector of atom $\alpha$,
\begin{eqnarray}
\ket{\psi^{i}} & = & \sum_{\mu}c_{\mu}^{i}\ket{\mu(\vec{r}-\vec{R_{\mu}})}\\
\ket{\psi^{j}} & = & \sum_{\nu}c_{\nu}^{j}\ket{\nu(\vec{r}-\vec{R_{\nu}})}
\end{eqnarray}
are Kohn-Sham molecular orbitals in the basis of atom-centered numerical
orbitals, and the transition charges are defined as 
\begin{equation}
q_{\alpha}^{ij}=\frac{1}{2}\sum_{\mu\in\alpha}\sum_{\nu}\left(c_{\mu}^{i}c_{\nu}^{j}S_{\mu\nu}+c_{\nu}^{i}c_{\mu}^{j}S_{\nu\mu}\right)
\end{equation}
This approximation fails if a transition happens between molecular
orbitals on the same atom, which are orthogonal by construction. In
this case the overlap matrix simplifies to the identity matrix and
the Mulliken approximation leads to vanishing transition charges.
Therefore the excitation energies for localized $n\to\pi^{*}$ excitations
are not shifted by the TD-DFT coupling matrix and no improvement of
these energies relative to the Kohn-Sham energies is achieved \cite{tddftb_extensions}.
Moreover, the oscillator strengths, which determine the shape of the
spectrum, can sometimes be sensitive to the way transition dipole
matrix elements are computed. For example, in reference \cite{tddftb_extensions}
an on-site correction to the dipole matrix element is introduced,
so that the $^{2}\Pi$ states of nitric oxide, which would be dark
using the Mulliken approximation, gain a small oscillator strength.

However, the oscillator strengths can also be calculated without approximation
from the transition dipoles between atomic orbitals which are assembled
using the Slater-Koster rules. The technical details are explained
in the appendix \ref{sec:dipoles_technical_details} (apparently this
approach has already been implemented in other DFTB codes, but is
not well documented in the literature).

\begin{figure}[h!]
\includegraphics[width=0.8\textwidth]{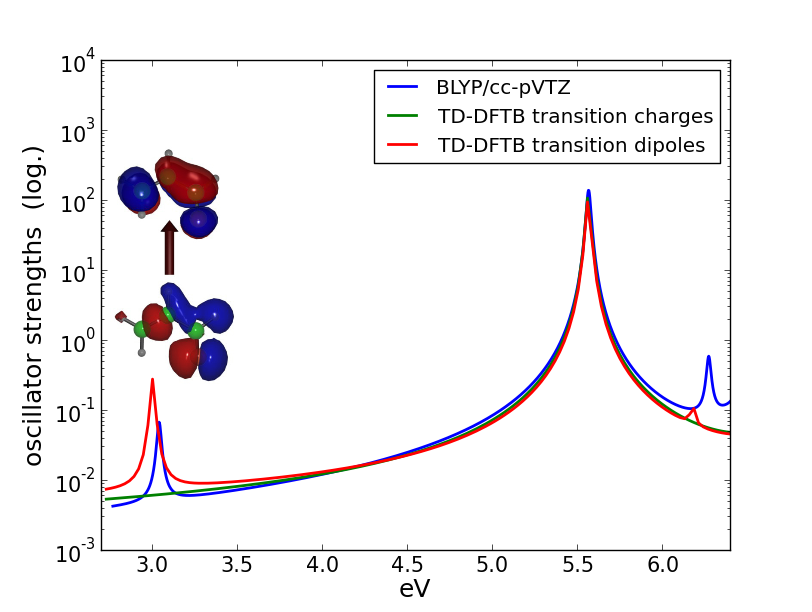}
\caption{\textbf{Absorption spectrum of acrolein}. The molecular structure
was optimized at the BLYP\cite{LYP,Becke_exact_exchange}/cc-pVTZ\cite{cc_pvtz}
level, and a range-separation parameter of $R_{\text{lr}}=20.0$ bohr
was used in the lc-TD-DFTB calculation. With the Mulliken approximation
(green curve) several states disappear, although they have a small
oscillator strength. For example, the lowest excited states at $\approx3$
eV is of $n\to\pi^{*}$ character and has a tiny oscillator strength
of $4\cdot10^{-4}$ which cannot be seen in the green curve.}

\label{fig:Mulliken_failure} 
\end{figure}

For transitions comprising orbitals on different atoms the oscillator
strengths derived from the Mulliken charges and from the tabulated
dipole matrix elements are very similar. However, when the transition
is confined to a single atom, the Mulliken approximation can miss
states which have weak oscillator strengths. As an example, consider
the acrolein molecule, where the $S_{1}$ state is characterized by
an excitation from the $n$ to the $\pi^{*}$ orbital on the oxygen.
Since the oxygen is much more electronegative than the carbon it is
bound to, both the $n$ and the $\pi^{*}$ orbitals consist largely
of the atomic oxygen orbitals. The oscillator strength is very small,
but it is not zero. Fig.\ref{fig:Mulliken_failure} compares the absorption
spectra when the oscillator strengths are calculated from Mulliken
charges or transition dipoles, respectively. 

\subsection{Diagnosing charge transfer states}

\label{sec:lambda_diagnostics} Visual inspection of the orbitals
involved in an excitation is normally required to characterize an
excited state as a charge transfer (CT) state. The authors of ref.
\cite{lambda_diagnostic} introduced a simple numeric test for detecting
problematic excitations. They quantified the degree of spatial overlap
between occupied and virtual orbitals of an excited state $I$ using
the quantity 
\begin{equation}
\Lambda=\sum_{o\in\text{occ}}\sum_{v\in\text{virt}}\left(C_{ov}^{I}\right)^{2}\int\vert\phi_{o}(\vec{r})\vert\vert\phi_{v}(\vec{r})\vert d^{3}r,\label{eqn:Lambda_original}
\end{equation}
where $C_{ov}^{I}$ is the coefficient for the single excitation from
the Kohn-Sham Slater determinant that replaces the occupied orbital
$o$ by the virtual orbital $v$ in the excited state $I$.

They found that the error of DFT functionals without long-range correction
such as PBE\cite{PBE} and B3LYP\cite{Becke_exact_exchange,B3LYP_assembly},
correlates with the value of $\Lambda$, which can vary between 0.0
and 1.0. Values below 0.5 indicate a charge transfer or Rydberg excitation,
for which PBE and B3LYP will probably be in large error.

Unfortunately, the integral over products of orbital moduli is difficult
to calculate in DFTB without resorting to numerical integration. Therefore
we replace the integral by 
\begin{equation}
O_{ov}=\int\vert\phi_{o}(\vec{r})\vert^{2}\vert\phi_{v}(\vec{r})\vert^{2}d^{3}r
\end{equation}
and define the new quantity 
\begin{equation}
\Lambda_{2}=\sum_{o\in\text{occ}}\sum_{v\in\text{virt}}\left(C_{ov}^{I}\right)^{2}\frac{O_{ov}}{\sqrt{O_{oo}O_{vv}}}\label{eqn:Lambda2}
\end{equation}
that should behave similarly to $\Lambda$ from eqn. \ref{eqn:Lambda_original}
. Applying again the monopole approximation, $O_{ov}$ can be approximated
by 
\begin{equation}
O_{ov}\approx\sum_{A}\sum_{B}q_{A}^{oo}q_{B}^{vv}\underbrace{\int F_{A}(\vert\vec{r}-\vec{R}_{A}\vert)F_{B}(\vert\vec{r}-\vec{R}_{B}\vert)d^{3}r}_{\Omega_{AB}}.
\end{equation}
The overlap integral $\Omega_{AB}$ of the spherical charge distributions
centered on atoms $A$ and $B$ can be performed analytically assuming
that $F$ has a Gaussian profile: 
\begin{equation}
F_{A}(\vec{r})=\frac{1}{\left(2\pi\sigma_{A}^{2}\right)^{3/2}}\exp\left(-\frac{\vert\vec{r}\vert^{2}}{2\sigma_{A}^{2}}\right)
\end{equation}
where the width of the distribution is inversely proportional to the
Hubbard parameter of atom A (see eqn. 29 of ref. \cite{DFTB_for_beginners}):
\begin{equation}
%\sigma_{A}=\frac{1.329}{\sqrt{8\ln(2)}}\frac{1}{U_{A}}.
\sigma_{A}=\frac{1}{\sqrt{\pi}}\frac{1}{U_{A}}.
\end{equation}
The integral is: 
\begin{equation}
\Omega_{AB}=\frac{1}{\left(2\pi\left(\sigma_{A}^{2}+\sigma_{B}^{2}\right)\right)^{3/2}}\exp\left(-\frac{1}{2}\frac{1}{\sigma_{A}^{2}+\sigma_{B}^{2}}\left(\vec{R}_{A}-\vec{R}_{B}\right)^{2}\right)
\end{equation}

For extended molecular systems the analysis of charge transfer by
looking at the orbitals can become very cumbersome as the grids on
which they are stored have to be very large. In the appendix \ref{sec:ct_analysis_drho}
we provide an alternative analysis method in terms of the difference
density between ground and excited stated, which is partitioned into
atomic contributions.

\section{Computational Details}

\label{sec:computational_details} The values for the confinement
radius $r_{0}$ and the Hubbard parameter $U_{H}$ that were used
to parametrize the electronic part of DFTB are shown in Table \ref{tbl:electronic_DFTB_parameters}.
The Kohn-Sham equations with the PW92\cite{PW92} local exchange-correlation
functional for the pseudo atoms were solved self-consistently using
the shooting method\cite{shooting_method}. In order to quickly assemble
the DFTB Hamiltonian and the transition dipoles, Slater-Koster tables
were generated for all pairwise combinations of valence orbitals for
the atoms H,C,N and O. The parameter for the long range correction
was set to $R_{\text{lr}}=3.03$ bohr (for comparison, CAM-B3LYP\cite{CAM_B3LYP}
uses $R_{\text{lr}}=\frac{1}{\mu}=3.03$ bohr, but only 65\% exact
exchange). So far we did not try to find an optimum value for $R_{\textbf{lr}}$.

\begin{table}[h!]
\centering %
\begin{tabular}{r|r|r}
Atom  & $r_{0}$ / bohr  & $U_{H}$ / Hartree \tabularnewline
\hline 
\hline 
H  & $1.084$  & $0.472$ \tabularnewline
C  & $2.657$  & $0.367$ \tabularnewline
N  & $2.482$  & $0.530$ \tabularnewline
O  & $2.307$  & $0.447$ %\\\hline
%  zn &  $4.900$  &  $0.267$  \\ 
\tabularnewline
\end{tabular}\caption{\textbf{Electronic DFTB parameters}. In general, the Hubbard parameter
is set to the difference between experimental ionization energy and
electron affinity, $U_{H}=$ IE\cite{ionization_energies}-EA\cite{electron_affinities},
where available, and the confinement radius is taken as $1.85\times$
the covalent radius as reported in \cite{covalent_radii}, $r_{0}=1.85r_{\text{cov}}$.
Deviating from this rule, the Hubbard parameter of nitrogen was set
to $0.530$ Hartree (taken from the DFTB implementation of \textsl{Hotbit}\cite{DFTB_for_beginners}),
since there is no experimental value for the electron affinity available
(the $N^{-}$ anion is unstable).}

\label{tbl:electronic_DFTB_parameters} 
\end{table}

The structures of the test molecules were taken from the database
\cite{helgaker_database} that has been published by the authors of
Ref. \cite{lambda_diagnostic}. The molecules N$_{2}$, CO, H$_{2}$CO
and HCl were excluded because the minimal basis set of occupied valence
orbitals used in DFTB is not suitable for describing Rydberg states.
In the calculations labeled as \textbf{lc-TD-DFTB} the long-range
correction as described above was included, Casida's equation \ref{eqn:Casidas_equation}
was used for solving the linear response equations and the transition
dipoles were assembled from Slater-Koster files. In the calculations
labeled as \textbf{TD-DFTB} the long-range correction was absent and
the transition dipoles were calculated using transition charges. For
comparison we also determined the excitation energies and oscillator
strengths with full TD-DFT using the CAM-B3LYP\cite{CAM_B3LYP} functional
and the cc-pVTZ\cite{cc_pvtz} basis set as implemented in the Gaussian
09\cite{g09} program package.

\section{Results}

\label{sec:results} Energies and oscillator strengths of specific
excitations computed with TD-DFTB, lc-TD-DFTB and CAM-B3LYP are compared
in Table \ref{tbl:comparison_table}. The HOMO-LUMO gaps are plotted in 
Fig. \ref{fig:HOMO_LUMO_gaps}.

\begin{figure}[h!]
\centering \includegraphics[width=0.6\textwidth]{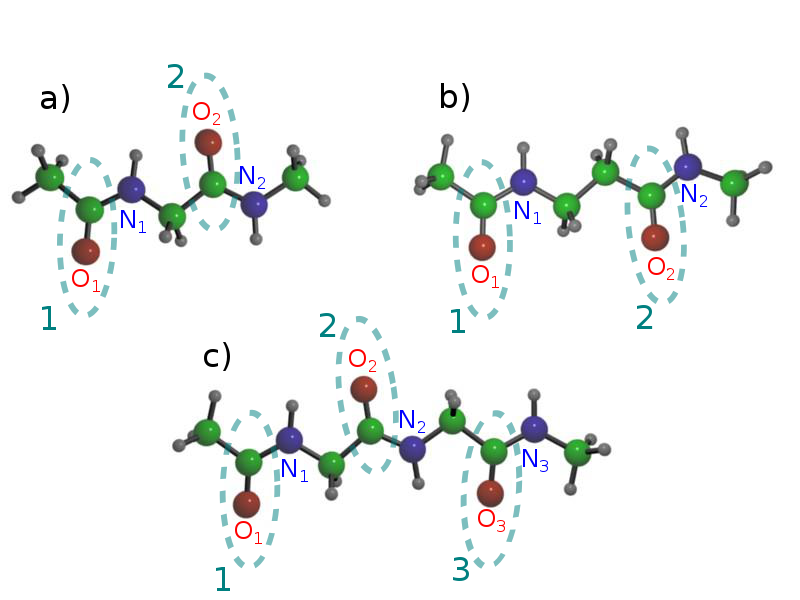}
\caption{\textbf{Peptide structures.} \textbf{a)} dipeptide, \textbf{b)} $\beta$-dipeptide and
\textbf{c)} tripeptide. The carbonyl groups used to classify the
excitations are encircled. }
\label{fig:peptide_structures} 
\end{figure}

\begin{figure}[h!]
\includegraphics[width=0.8\textwidth]{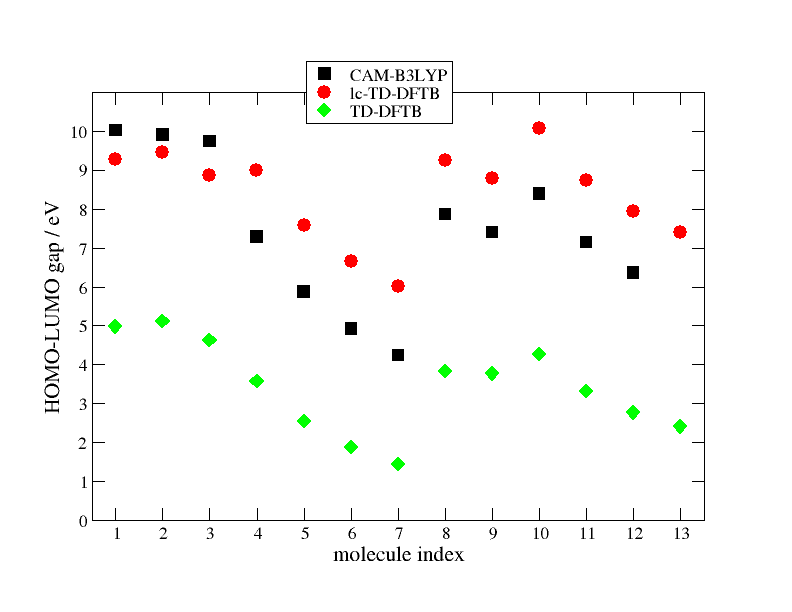}
\caption{\textbf{HOMO-LUMO gaps} for the molecules from the test set. The names of the molecules encoded by the numbers on the x-axis can be found in the first column of table \ref{tbl:comparison_table}. As expected from a long-range corrected functional the HOMO-LUMO gaps are larger for lc-TD-DFTB than for TD-DFTB.}
\label{fig:HOMO_LUMO_gaps}
\end{figure}

\textbf{Peptides}. The first molecules in the test set are three model
peptides of increasing length, a dipeptide, a $\beta$-dipeptide and
a tripeptide. The labels $n(O_i)$, $\pi(N_i)$ and $\pi_i^{*}$, which
are used to classify the excited states, refer to the lone electron pair
on the the $i$-th oxygen, the binding $\pi$-orbital containing the $i$-th nitrogen
and the anti-bonding $\pi$-orbital on the $i$-th carbonyl group. The ordering of the carbonyl
groups and atoms is depicted in Fig. \ref{fig:peptide_structures}.

The energies of the lowest localized excitations are well reproduced
both by TD-DFTB and lc-TD-DFTB. The energies of charge transfer excitations
are underestimated, although lc-TD-DFTB puts them much closer to the CAM-B3LYP
results and correctly predicts that the band of charge transfer states
should be located above the local excitations. As opposed to this,
TD-DFTB produces charge transfer states that lie below the lowest
local excitation. This problem gets worse as the peptide grows: In
the tripeptide TD-DFTB erroneously predicts the lowest excitation
to be a long-range charge transfer from $n(O_1)$ to $\pi_{3}^{*}$
with an energy of $4.59$ eV, whereas the CAM-B3LYP energy of this
excitation would be $8.68$ eV.

\textbf{Acenes}. The next molecules are the smallest 4 polyacenes
Naphthalene (n=1), Anthracene (n=2), Tetracene (n=3) and Pentacene
(n=4). The large degree of conjugation leads to orbitals that are delocalized
over the entire molecule. TD-DFTB underestimates the energies of all states
consistently by $< 0.5$ eV, and this error remains stable with the size of the acenes.
With lc-TD-DFTB the $B_{2u}$ states deviate no more than $0.1$ eV from the CAM-B3LYP reference
values. 

\textbf{N-phenylpyrrole.} In this heterocyclic aromatic compound,
TD-DFTB underestimates local excitations by $\approx 1$ eV while
lc-TD-DFTB is correct to within $0.1$ eV. 
Without long-range correction the lowest state with $A_1$ symmetry has charge transfer character ($\Lambda_2 = 0.02$). The long-range correction shifts this state
to higher energies so that the lowest $A_1$ state now belongs to a local excitation ($\Lambda_2=0.49$), as it should.

The second $B_{2}$ and $A_{1}$ states, which involve an electron transfer from the pyrrole
ring to the benzene ring, are predicted far too low in energy by TD-DFTB.

\textbf{DMABN}. 4-(N,N-dimethylamino)benzonitrile (DMABN) possesses
a low-lying charge transfer state that is formed when the nitrogen
on one side of the phenyl ring donates charge to the -C$\equiv$N
group on the opposite side. 
Although the ordering of the states is correct even without long-range correction,
lc-TD-DFTB comes much closer to the reference
values than plain TD-DFTB.

\textbf{Polyacetylenes}. As with the acenes, TD-DFTB energies are too low by $0.5$ eV
as compared to the lc-TD-DFTB and CAM-B3LYP values.

% TABLE

\begin{table}[h!]
\centering %
\begin{tabular}{clclcccc}
& &  &  &  &  &  & \tabularnewline
\hline 
\hline 
& Molecule  & State  & $\Lambda_{2}$  & Type  & TD-DFTB  & lc-TD-DFTB  & CAM-B3LYP \tabularnewline
\hline 
1 & Dipeptide          & A'' $n(O_1) \to \pi_1^*$    & 0.61 & L  & 5.48 (0.000) & 5.33 (0.000) & 5.68 (0.001) \tabularnewline
  &                    & A'' $n(O_2) \to \pi_2^*$    & 0.75 & L  & 5.44 (0.000) & 5.40 (0.000) & 5.92 (0.001) \tabularnewline
  &                    & A'  $\pi(N_1) \to \pi_2^*$  & 0.08 & CT & 5.54 (0.002) & 6.23 (0.015) & 7.00 (0.010) \tabularnewline
  &                    & A'' $n(O_1) \to \pi_2^*$    & 0.20 & CT & 4.92 (0.000) & 5.98 (0.000) & 7.84 (0.000) \tabularnewline
2 & $\beta$-dipeptide  & A'' $n(O_1) \to \pi_1^*$    & 0.77 & L  & 5.47 (0.000) & 5.39 (0.000) & 5.67 (0.001) \tabularnewline
  &                    & A'' $n(O_2) \to \pi_2^*$    & 0.77 & L  & 5.45 (0.000) & 5.39 (0.000) & 5.76 (0.000) \tabularnewline
  &                    & A'  $\pi(N_1) \to \pi_2^*$  & 0.56 & CT & 5.64 (0.000) & 7.37 (0.558) & 7.42 (0.328) \tabularnewline
  &                    & A'' $n(O_1) \to \pi_2^*$    & 0.01 & CT & 5.05 (0.000) & 6.71 (0.000) & 8.38 (0.008) \tabularnewline
3 & Tripeptide         & A'' $n(O_1) \to \pi_1^*$    & 0.59 & L  & 5.51 (0.000) & 5.34 (0.000) & 5.72 (0.001) \tabularnewline
  &                    & A'' $n(O_2) \to \pi_2^*$    & 0.58 & L  & 5.50 (0.000) & 5.37 (0.000) & 5.93 (0.001) \tabularnewline
  &                    & A'' $n(O_3) \to \pi_3^*$    & 0.72 & L  & 5.44 (0.000) & 5.43 (0.000) & 6.00 (0.001) \tabularnewline
  &                    & A'  $\pi(N_1) \to \pi_2^*$  & 0.07 & CT & 5.56 (0.002) & 6.25 (0.014) & 6.98 (0.014) \tabularnewline
  &                    & A'  $\pi(N_2) \to \pi_3^*$  & 0.16 & CT & 5.78 (0.002) & 6.52 (0.032) & 7.68 (0.102) \tabularnewline
  &                    & A'' $n(O_1) \to \pi_2^*$    & 0.19 & CT & 4.93 (0.000) & 5.99 (0.000) & 7.78 (0.000) \tabularnewline
  &                    & A'' $n(O_2) \to \pi_3^*$    & 0.28 & CT & 5.16 (0.000) & 6.33 (0.000) & 8.25 (0.000) \tabularnewline
  &                    & A'  $\pi(N_1) \to \pi_3^*$  & 0.03 & CT & 5.20 (0.000) & 8.35 (0.000) & 8.51 (0.007) \tabularnewline
  &                    & A'' $n(O_1) \to \pi_3^*$    & 0.01 & CT & 4.59 (0.000) & 9.04 (0.000) & 8.68 (0.000) \tabularnewline
\hline 
4 & Acene (n=1)  & $B_{2u}$  & 0.83  & DL  & 4.27 (0.007)  & 4.53 (0.006)  & 4.62 (0.000) \tabularnewline
  &              & $B_{1u}$  & 0.77  & DL  & 4.02 (0.044)  & 4.84 (0.046)  & 4.67 (0.071) \tabularnewline
5 & Acene (n=2)  & $B_{1u}$  & 0.77  & DL  & 3.00 (0.047)  & 3.84 (0.067)  & 3.53 (0.076) \tabularnewline
  &              & $B_{2u}$  & 0.83  & DL  & 3.66 (0.026)  & 4.02 (0.021)  & 4.04 (0.001) \tabularnewline
6 & Acene (n=3)  & $B_{1u}$  & 0.74  & DL  & 2.32 (0.040)  & 3.19 (0.073)  & 2.76 (0.071) \tabularnewline
  &              & $B_{2u}$  & 0.82  & DL  & 3.27 (0.056)  & 3.69 (0.047)  & 3.65 (0.003) \tabularnewline
7 & Acene (n=4)  & $B_{1u}$  & 0.72  & DL  & 1.85 (0.033)  & 2.74 (0.076)  & 2.22 (0.064) \tabularnewline
  &              & $B_{2u}$  & 0.81  & DL  & 3.01 (0.100)  & 3.48 (0.087)  & 3.39 (0.008) \tabularnewline
\hline 
8 & N-phenylpyrrole  & $B_{2}$  & 0.48  & L  & 3.96 (0.005)  & 4.98 (0.003)  & 5.06 (0.013) \tabularnewline
  &                  & $A_{1}$  & 0.49  & L  & 3.99 (0.000)  & 5.07 (0.458)  & 5.12 (0.365) \tabularnewline
  &                  & $B_{2}$  & 0.29  & CT & 4.30 (0.009)  & 5.24 (0.015)  & 5.27 (0.015) \tabularnewline
  &                  & $A_{1}$  & 0.02  & CT & 4.51 (0.352)  & 6.18 (0.000)  & 5.92 (0.179) \tabularnewline
\hline 
9 & DMABN        & $B$       & 0.37  & L   & 4.27 (0.023)  & 4.47 (0.022)  & 4.72 (0.024) \tabularnewline
  &              & $A$       & 0.49  & CT  & 4.52 (0.308)  & 4.95 (0.453)  & 4.91 (0.520) \tabularnewline
\hline 
10 & Polyacetylene (n=2)  & $B_{u}$  & 0.68  & DL  & 5.57 (0.500)  & 6.21 (0.420)  & 6.04 (0.706) \tabularnewline
11 & Polyacetylene (n=3)  & $B_{u}$  & 0.69  & DL  & 4.54 (0.813)  & 5.10 (0.707)  & 5.03 (1.110) \tabularnewline
12 & Polyacetylene (n=4)  & $B_{u}$  & 0.69  & DL  & 3.88 (1.133)  & 4.43 (0.995)  & 4.39 (1.533) \tabularnewline
13 & Polyacetylene (n=5)  & $B_{u}$  & 0.69  & DL  & 3.41 (1.437)  & 3.98 (1.269)  & 3.94 (1.961) \tabularnewline
\hline 
\end{tabular}\caption{TD-DFTB and lc-TD-DFTB excitation energies and reference values from
TD-DFT calculations at the CAM-B3LYP/cc-pVTZ level. Energies are in
eV and oscillator strengths are given in brackets. The type of excitation
(L = local, CT = charge transfer, DL = delocalized) was assigned by
inspecting the dominant pair of occupied and virtual orbitals in the
transitions. The $\Lambda_{2}$ values are calculated for lc-TD-DFTB.}

\label{tbl:comparison_table} 
\end{table}

\subsection{Correlation between errors and $\Lambda_{2}$}

\begin{figure}[h!]
\includegraphics[width=0.8\textwidth]{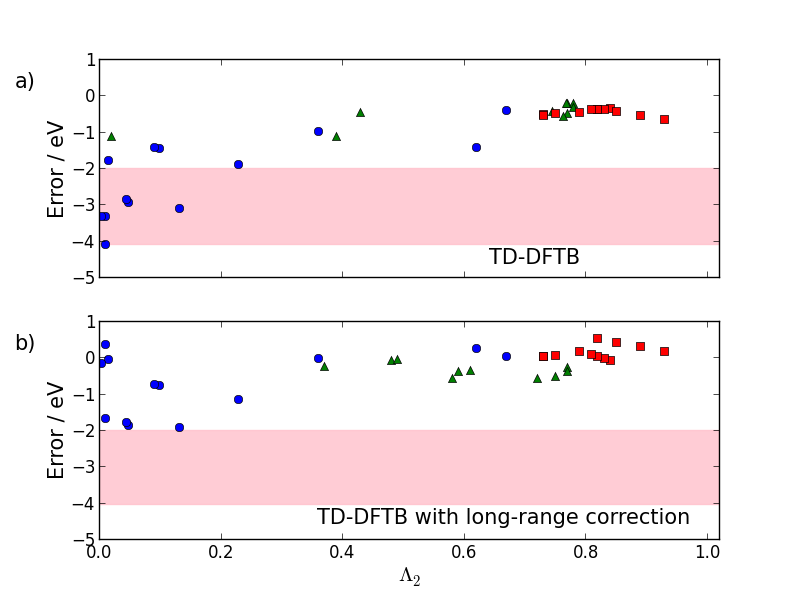}
\caption{Deviation of excitation energies from CAM-B3LYP reference values for
\textbf{a)} TD-DFTB and \textbf{b)} lc-TD-DFTB plotted against $\Lambda_{2}$
(a measure of spatial overlap defined in eqn. \ref{eqn:Lambda2}),
local excitations ($\blacktriangle$), charge transfer excitations
($\bullet$), delocalized excitations ({\footnotesize{$\blacksquare$}}).
The area into which errors from charge transfer states fall without
long-range correction is highlighted in pink.}

\label{fig:deviation_against_lambda2} 
\end{figure}

When the deviations of the excitation energies relative to the CAM-B3LYP
values are plotted against the degree of spatial overlap $\Lambda_{2}$
(see Fig. \ref{fig:deviation_against_lambda2}), a clear correlation
is visible. $\Lambda_{2}$ values close to 0.0 can be associated with
charge transfer states, values around 0.5 with local excitations and
values close to 1.0 with strongly delocalized excitations. For charge
transfer states, the long range correction reduces the maximum error
from -4.0 to -2.0 eV. For local excitations, TD-DFTB and lc-TD-DFTB
have similar errors - TD-DFTB underestimates energies by at most 0.5
eV, while lc-TD-DFTB overestimates them by the same amount. For strongly
delocalized excitations, as they occur in the acene series, lc-TD-DFTB
shifts the error to the positive region with respect to TD-DFTB, and
reduces somewhat the absolute values of the error.

Table \ref{tbl:mean_errors} gives the mean errors averaged over the
whole set of test molecules. In summary, energies of localized, charge
transfer as well as delocalized states are systematically improved
by lc-TD-DFTB. In particular, the long range correction particularly
markedly the energies of delocalized and charge transfer states while
the energies of the localized states are only marginally better.

\begin{table}[h!]
\begin{tabular}{ccccc}
Method  & Total  & Local  & Charge Transfer  & Delocalized \tabularnewline
\hline 
TD-DFTB  & 1.13  & 0.51  & 2.22  & 0.46 \tabularnewline
lc-TD-DFTB  & 0.46  & 0.34  & 0.83  & 0.17 \tabularnewline
\end{tabular}\caption{\textbf{Mean errors} (in eV) relative to the CAM-B3LYP excitation
energies for the molecules in the test set. The long-range correction
particularly improves energies of delocalized and charge transfer
states. }

\label{tbl:mean_errors} 
\end{table}

\FloatBarrier

\subsection{Are potential energy surfaces with lc-TD-DFTB accurate enough?}

The good agreement of lc-TD-DFTB absorption spectra with
those obtained from full long-range corrected TD-DFT around equilibrium
geometries unfortunately does not guarantee that global features of the excited
potential energy surfaces are equally well reproduced.

\begin{figure}[h!]
\includegraphics[width=0.8\textwidth]{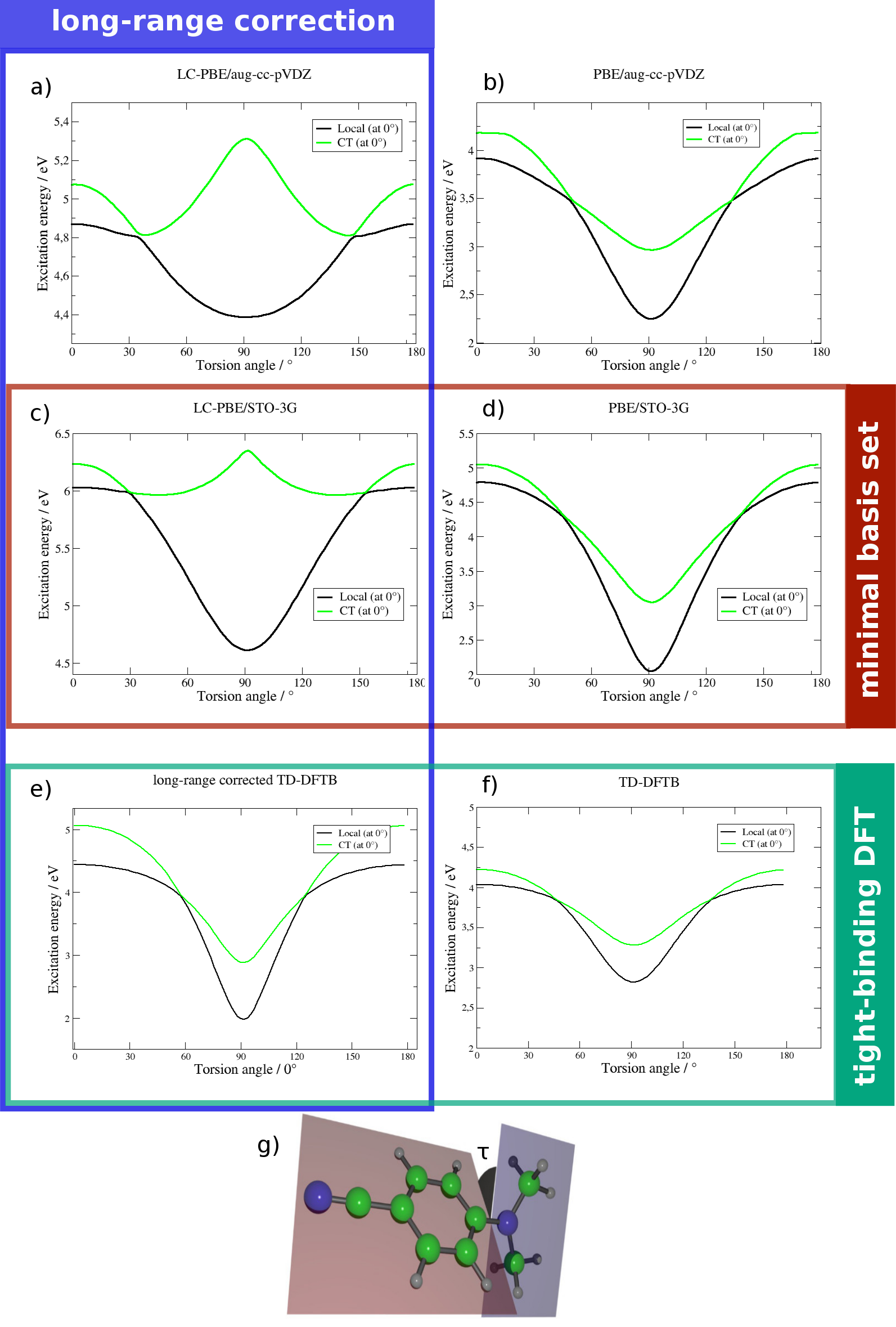} \caption{Excitation energy profile of DMABN along torsion angle. \textbf{a)},\textbf{b)},\textbf{c)}
and \textbf{d)} with full TD-DFT using the PBE functional; \textbf{e)}
and \textbf{f)} with tight-binding DFT.}

\label{fig:DMABN-torsion_angle} 
\end{figure}

\begin{figure}[h!]
\includegraphics[width=0.8\textwidth]{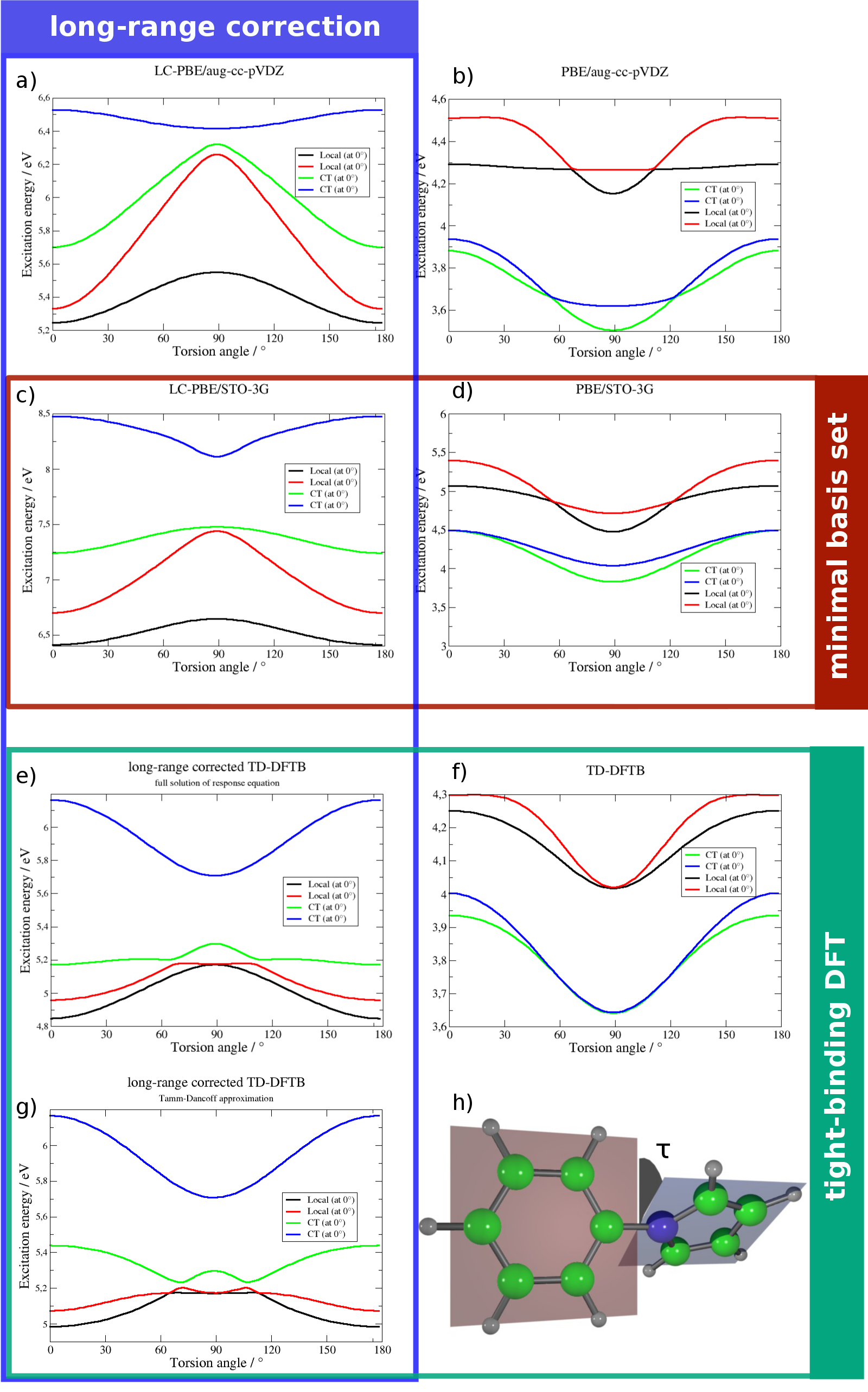}
\caption{Excitation energy profile of N-phenylpyrrole along torsion angle.
\textbf{a)},\textbf{b)},\textbf{c)} and \textbf{d)} with full TD-DFT
using the PBE function; \textbf{e)},\textbf{f)} and \textbf{g)} with
tight-binding DFT. In \textbf{e)} the lc-TD-DFTB spectrum has been
obtained from the full solution of eqn. \ref{eqn:hermitian_eigval}
while in \textbf{g)} the Tamm-Dancoff approximation was employed.}

\label{fig:N-phenylpyrrole_torsion_angle} 
\end{figure}

We have investigated the accuracy of the excited state potentials by exploring profiles of the excitation energies along the torsion
angles in two of the molecules from the test set, DMABN and N-phenylpyrrole
(see insets in Figs. \ref{fig:DMABN-torsion_angle} and \ref{fig:N-phenylpyrrole_torsion_angle}
for the definition of the torsion angle $\tau$), and tried to identify
the approximations in TD-DFTB that lead to the discrepancies with
respect to full TD-DFT.

Both molecules show dual fluorescence\cite{dual_florescence_DMABN,dual_florescence_N-PP}
since the excited charge transfer state has an energy minimum at a
dihedral angle of 90$^{\circ}$, while the ground state is more or
less planar. As shown for DMABN\cite{Tozer_DMABN}, density functionals
without long-range correction such as PBE or B3LYP fail to predict
an energy maximum of the charge transfer state at the twisted geometry.
Only long-range corrected functionals are faithful to the shape of
the potential energy surface.

%%%
In order to separate the influence of the functional from the influence
of the basis size, we first compare the torsion profiles for the PBE
\cite{PBE} with those of the LC-PBE functional \cite{LC_Gaussian}
which is the long-range corrected version of PBE - both with the aug-cc-pVDZ\cite{cc_pvtz}
basis set (Fig.\ref{fig:DMABN-torsion_angle}\textbf{a)} vs. \textbf{b)}
and Fig.\ref{fig:N-phenylpyrrole_torsion_angle}\textbf{a)} vs. \textbf{b)}).
In the next step we reduce the basis to a minimal one (STO-3G\cite{STO-3G})
(Fig.\ref{fig:DMABN-torsion_angle}\textbf{c)} vs. \textbf{d)} and
Fig.\ref{fig:N-phenylpyrrole_torsion_angle}\textbf{c)} vs. \textbf{d)}).
Clearly, the absence of a long-range exchange term changes the profiles
qualitatively, whereas the reduction of the basis size does not -
except for a large energy increase of all excited states. Now, the
stage is set for comparing with tight-binding DFT.

%%%

In standard TD-DFTB, the charge transfer states collapse to a minimum
at $\tau=90^{\circ}$ for both molecules as expected of a functional
lacking long-range exchange (Fig.\ref{fig:DMABN-torsion_angle}\textbf{f)}
and Fig.\ref{fig:N-phenylpyrrole_torsion_angle}\textbf{f)}). Density
functionals with local exchange cannot describe charge transfer over
long distances because the transition density $\phi_{o}(\vec{r})\phi_{v}(\vec{r})$
vanishes if the occupied and virtual orbital are spatially separated\cite{CT_non_local_x}.

For lc-TD-DFTB, one would expect an improved description of the charge
transfer states. While the energy maximum at the twisted geometry
is at least qualitatively reproduced for N-phenylpyrrole (Fig.\ref{fig:N-phenylpyrrole_torsion_angle}\textbf{e)},\textbf{g)}),
for DMABN lc-TD-DFTB performs even worse than TD-DFTB (Fig.\ref{fig:DMABN-torsion_angle}\textbf{e)}):
The lc-TD-DFTB torsion profile shows an even more pronounced dip at
90$^{\circ}$, but for a different reason. This becomes evident from Tables \ref{tbl:coupling_matrix_DMABN_0}
and \ref{tbl:coupling_matrix_DMABN_90} which show the blocks of the long-range
part of the coupling matrix $K_{lr}$ that mixes different transitions
between frontier orbitals. The lowest two excited states in DMABN
are dominated by excitations from the HOMO-1 or HOMO into the LUMO
or LUMO+1. A small off-diagonal matrix element of $-0.01$ is present
at $\tau=0^{\circ}$, whereas at $\tau=90^{\circ}$ the block becomes
almost diagonal (the largest off-diagonal element is below smaller
than 0.0001). Therefore at the twisted geometry, the coupling matrix
cannot correct the excitation energies. The vanishing of the long-range
coupling is probably due to the Mulliken approximation that reduces
the transition density to spherically symmetric charge distributions
around atoms.

\begin{table}[h!]
\begin{tabular}{c|ccc}
$K_{lr}$  & H-1 $\to$ L  & H $\to$ L  & H $\to$ L+1 \tabularnewline
\hline 
%------------------------------------------------
H-1 $\to$ L  & 0.1798  & 0.0000  & -0.0106 \tabularnewline
H $\to$ L  & 0.0000  & 0.1543  & -0.0000 \tabularnewline
H $\to$ L+1  & -0.0106  & -0.0000  & 0.1721 \tabularnewline
\end{tabular}\caption{long-range coupling matrix $K_{lr}$ for the frontier orbitals at
$\tau=0^{\circ}$ for DMABN.}

\label{tbl:coupling_matrix_DMABN_0} 
\end{table}

\begin{table}[h!]
\begin{tabular}{c|ccc}
$K_{lr}$  & H-1 $\to$ L  & H $\to$ L  & H $\to$ L+1 \tabularnewline
\hline 
%------------------------------------------------
H-1 $\to$ L  & 0.1520  & -0.0001  & -0.0001 \tabularnewline
H $\to$ L  & -0.0001  & 0.1590  & -0.0000 \tabularnewline
H $\to$ L+1  & -0.0001  & -0.0000  & 0.1690 \tabularnewline
\end{tabular}\caption{long-range coupling matrix $K_{lr}$ at $\tau=90^{\circ}$ for DMABN.}

\label{tbl:coupling_matrix_DMABN_90} 
\end{table}

Comparing the torsion profile of N-phenylpyrrole for the aug-cc-pVDZ
and the minimal STO-3G basis sets allows one to identify a second
source of error. Diffuse orbitals probably play a large role in charge
transfer states, where orbitals participating in a transition overlap
only slightly with each other's fuzzy tails. The torsion profile at
the LC-PBE/STO-3G level (Fig. \ref{fig:N-phenylpyrrole_torsion_angle}
\textbf{c)}) is comparable to lc-TD-DFTB (Fig. \ref{fig:N-phenylpyrrole_torsion_angle}
\textbf{e)} and \textbf{d)}), which also uses a minimal set of valence
orbitals. In both cases the maxima of the lowest 3 excited states
are less pronounced than in the LC-PBE/aug-cc-pVDZ (Fig. \ref{fig:N-phenylpyrrole_torsion_angle}
\textbf{a)}) case.

\FloatBarrier

\section{Conclusion and Outlook}

Tight-binding DFT is a very efficient method for calculating the electronic
structure of large organic molecules around their equilibrium structures.
This efficiency rests on the precalculation of matrix elements using
the Slater-Koster rules and on approximations for the two-center integrals
using Mulliken charges. Any improvement of the method should not deviate
from these principles as it would forfeit speed of execution. We have
presented two improvements that add very little computational overhead,
the long-range correction in the spirit of CAM-B3LYP and the calculation
of oscillator strengths from Slater-Koster files for transition dipoles.
In full TD-DFT, the computational cost increases considerably if range-separated
functionals are used, not so in the tight-binding approximation.

Two measures of charge transfer were tested which can be extracted
easily from TD-DFTB calculations, the quantity $\Lambda_{2}$ and
the particle-hole separation. 
These can be used to detect cases of problematic
behaviour involving long charge-transfer character of excited
states. Application of the method to a test set of molecules with
problematic charge transfer states led to an average error of 0.3
eV for local excitations, 0.8 eV for charge transfer excitations and
0.2 eV for delocalized excitations. Especially energies of charge
transfer states are improved by over 1.0 eV relative to tight-binding
DFT without range-separation.
A more general test of the method for a set of 
common organic molecules\cite{thiel_set} is compiled
in the supporting information.

Of course, being a semi-empirical method, the errors of lc-TD-DFTB
also depend on the right choice of transferable parameters and more
work needs to focus on optimizing these parameters, in particular
the range-separation radius $R_{\text{lr}}$ and the choice of the 
switching function (see appendix \ref{sec:appendix_switching_functions}. 
For simplicity, we derived
the electronic parameters (the confinement radius $r_{0}$ and the
Hubbard parameter $U_{H}$) from experimental data. Previous work
on the parametrization of DFTB\cite{parametrization_dftb3,parametrization_heine,parametrization_zinc}
has proven that, by fitting the parameters to full DFT calculations
on training sets of representative molecules, the accuracy of ab-initio
electronic structure calculations can be reached.

In the future, we intend to use lc-TD-DFTB to study polymers containing e. g. organic 
chromophores. In this context, a comparison with other semi-empirical
methods is pertinent. While some approximations are similar to semi-empirical
wavefunction based methods, the derivation from time-dependent density
functional theory is advantageous from the point of view of computational
efficiency since only single excitations are included. This approximation 
is often sufficient 
for describing low-lying bright excited states.
The problem that the active space grows quickly with the system size
becomes relevant only for much larger systems, when the coupling matrix
cannot be diagonalized with all single excitations included. Therefore
the method is suited for predicting absorption spectra of large systems
with a few thousand second row atoms.

\section{Acknowledgements}

The authors would like to thank Prof. M. E. Casida for helpful comments
on an early draft of this paper. A. H. and R. M. acknowledge funding
by the Deutsche Forschungsgemeinschaft (DFG SPP 1391 Ultrafast Nanooptics
MI-1236/3-2). R. M. acknowledges the support by the ERC Consolidator Grant
DYNAMO (Grant Nr. 646737)

\FloatBarrier

\appendix
%dummy comment inserted by tex2lyx to ensure that this paragraph is not empty

\section{Different approximations for the $\gamma_{AB}$ matrix}
\label{sec:appendix_gamma_matrix}
Reproducing DFTB results requires some care not only 
due to the different parametrizations but 
also because the various implementations employ slightly different approximations. One such ambiguity concerns 
the definition of the $\gamma$-matrix:
\begin{equation}
\gamma_{AB} = \int \int F_A(\vert \vec{r}_1 - \vec{R}_A \vert) \frac{1}{\vert \vec{r}_1 - \vec{r}_2 \vert} F_B(\vert \vec{r}_2 - \vec{R}_B \vert) d^3r_1 d^3r_2
\end{equation}
where $R = \vert \vec{R}_A - \vec{R}_B \vert$.

 The $\gamma$-matrix describes the change of the Coulomb energy due to charge redistribution between the atoms $A$ and $B$.
 Charge fluctuations are assumed to be spherically symmetric around each atom but the exact functional form is unknown.
 Slater and Gaussian functions are obvious candidates, since the Coulomb integrals are well-known for these functions.
 In some DFTB-implementations, the $\gamma$-matrix is based on Slater functions (presumably DFTB+ and the older code by Seifert)
 while in others it is based on Gaussian functions (Hotbit\cite{DFTB_for_beginners}).
\subsection{Slater functions}
If the charge fluctuation around an atomic center $A$ is modelled by a Slater function,
\begin{equation}
F_A(\vert \vec{r} - \vec{R}_A \vert) = \frac{\tau_A^3}{8 \pi} \exp\left(-\tau_A \vert \vec{r} - \vec{R}_A \vert \right)
\end{equation}
the Coulomb integral between two such charge distributions separated by a distance $R=\vert \vec{R}_A - \vec{R}_B \vert$ reads:
\begin{equation}
\begin{split}
\gamma_{AB} = \frac{1}{R} \Big[1 &+ \frac{\tau_B^4 \left(\tau_B^2 (2+\tau_A R) - \tau_A^2 (6+\tau_A R)\right)}{2 \left(\tau_A^2 - \tau_B^2\right)^3} e^{-\tau_A R} \\
  &- \frac{\tau_A^4 \left(\tau_A^2 (2+\tau_B R) - \tau_B^2 (6+\tau_B R)\right)}{2 \left(\tau_A^2 - \tau_B^2\right)^3} e^{-\tau_B R}   \Big]
\end{split}
\end{equation}
% tau is fixed by the requirement that gAA(R=0) = UA
In the limit, that the charge distributions are centered at the same position, the $\gamma$-matrix becomes:
\begin{equation}
\lim_{R \to 0} \gamma_{AB} = \frac{\tau_A \tau_B \left(\tau_A^2 + 3 \tau_A \tau_B + \tau_B^2 \right)}{2 \left(\tau_A^2 + \tau_B^2\right)^3}
\end{equation}

The width parameter $\tau_A$ is fixed by the requirement $\gamma_{AA}(R=0) = U_A$:
\begin{equation}
\tau_A = \frac{16}{5} U_A
\end{equation}

If both atoms are of the same type, one finds:
\begin{equation}
\lim_{\tau_A \to \tau_B = \tau} \gamma_{AB}(R) = \frac{1}{R} \left[1 - \frac{48 + 33 (\tau R) + 9 (\tau R)^2 + (\tau R)^3}{48} e^{-\tau R} \right]
\end{equation}

\subsection{Gaussian functions}
% cite hotbit
In the case of Gaussian functions, the charge distribution is modelled by 
\begin{equation}
F_A(\vert \vec{r} - \vec{R}_A \vert) = \frac{1}{\left(2 \pi \sigma_A^2\right)^{3/2}} \exp\left(-\frac{(\vec{r}-\vec{R}_A)^2}{2 \sigma_A^2} \right)
\end{equation}
Again, the Coulomb integral between two Gaussian charge fluctuations can be performed analytically, yielding:
% analytical integral
\begin{equation}
\gamma_{AB} = \frac{\erf(C_{AB} R)}{R}
\end{equation}
with 
\begin{equation}
C_{AB} = \frac{1}{\sqrt{2 (\sigma_A^2 + \sigma_B^2)}}
\end{equation}
$\sigma_A$ is determined by
\begin{equation}
\lim_{R \to 0} \gamma_{AA} = \lim_{R \to 0} \frac{\erf\left(\frac{R}{2 \sigma_A} \right)}{R} = \frac{1}{\sqrt{\pi} \sigma_A} \stackrel{!}{=} U_A
\end{equation}
which leads to
\begin{equation}
\sigma_A = \frac{1}{\sqrt{\pi} U_A}
\end{equation}
for the width parameter.

\section{Long-range switching functions and the $R_{AB} \to 0$ limit}
\label{sec:appendix_switching_functions}
Our long-range correction for DFTB has been criticized for neglecting the short range correction to the exchange energy and using an approximate $\gamma^{lr}_{AB}$ that has the supposedly wrong $R_{AB} \to 0$ limit. Both issues are related and we would like to defend our method by the following argument:

The error function is not the only and probably not the best way to separate the Coulomb potential into a short and a long range part.
For a general switching function $f(r)$ the separation becomes:
\begin{equation}
\frac{1}{r} = \underbrace{\frac{1-f(r)}{r}}_{\text{short range}} + \underbrace{\frac{f(r)}{r}}_{\text{long range}}
\end{equation}

Iikura's LC-functionals \cite{LC_Gaussian} and Niehaus' range-separated DFTB employ the error function
\begin{equation}
f_{\text{erf}}(r) = \erf(\omega r) \quad \quad \text{ with }  \omega = \frac{1}{R_{\text{lr}}}.
\end{equation}

But there are other choices. Toulouse \cite{LC_Savin} has shown that adding a Gaussian to the switching function (termed \textsl{erfgau})
\begin{equation}
f_{\text{erfgau}}(r) = \erf(\omega r) - \frac{2}{\sqrt{\pi}} (\omega r) \exp\left(-\frac{1}{3} (\omega r)^2 \right)
\end{equation}
has the advantage that the correction to the exchange functional that needs to be added to compensate the presence of the long-range HF exchange at short distance is reduced. An expansion of the exchange functional around $\omega = 0$,
\begin{equation}
E^{sr}_{x,\text{erfgau}} = E_x + O(\omega^5),
\end{equation}
demonstrates that the correction is negligible for $\omega < 1$. 

The limit of the long-range $\gamma$-matrix for $R_{AB} \to 0$ depends also on the switching function. We approximate this matrix as:
\begin{equation}
\begin{split}
\gamma_{AB}^{\text{lr}} &= \int \int F_A(\vert \vec{r}_1 - \vec{R}_A \vert) \frac{f(r_{12})}{r_{12}} F_B(\vert \vec{r}_2 - \vec{R}_B \vert) d^3r_1 d^3r_2 \\
 &\approx f(R_{AB}) \int \int F_A(\vert \vec{r}_1 - \vec{R}_A \vert) \frac{1}{r_{12}} F_B(\vert \vec{r}_2 - \vec{R}_B \vert) d^3r_1 d^3r_2 \\
 &= f(R_{AB}) \gamma_{AB} \label{eqn:gamma_lr_approx}
\end{split}
\end{equation}

One can argue that the approximation is not warranted, since for \textsl{erf} the numerical solution of the integral leads to a maximum for $R_{AB} \to 0$ (see Fig. 1 in reference \cite{range_separated_DFTB}), while our approximation gives $0$. However, if \textsl{erfgau} is used as a switching function the integral approaches a minimum in the limit $R_{AB} \to 0$. The shape of $\gamma^{lr}(R_{AB})$ at small distance depends strongly on the choice of the switching function, while the large distance limit is not affected.

We proceed by computing $\gamma^{lr}$ numerically for the \textsl{erf} and \textsl{erfgau} switching functions and compare the numerically exact results with the approximation in eqn. \ref{eqn:gamma_lr_approx}. It should become clear that the short range behaviour is immaterial and that it is actually beneficial that the long-range correction vanishes for $R_{AB} \to 0$, as then the short-range modifications to the exchange functional (or the error incurred by neglecting them) are smaller. 

Assuming a spherical distribution
\begin{equation}
F_A\left(\vert \vec{r} - \vec{R}_A \vert \right) = \frac{\tau_A^2}{8 \pi} \exp\left(-\tau_A \vert \vec{r} - \vec{R}_A \vert\right)
\end{equation}

the integral can be transformed into a one-dimensional integral (see Niehaus \cite{range_separated_DFTB})
\begin{equation}
\gamma_{AB}^{\text{lr}} = \frac{\tau_A^4 \tau_B^4}{\pi R_{AB}} \frac{1}{\imath} \int_{0}^{\infty} \frac{\sin\left(q R_{AB}\right)}{\left(q^2 + \tau_A^2\right)^2 \left(q^2 + \tau_B^2\right)^2} \tilde{f}(q) dq \label{eqn:gamma_lr_exact}
\end{equation}
where $\tilde{f}(q)$ is the Fourier transform of the switching function:
\begin{equation}
\tilde{f}(q) = \int_{-\infty}^{\infty} f(r) e^{\imath q r} dr \quad \quad \text{ assuming } f(-r) = -f(r)
\end{equation}
The Fourier transforms of the two switching functions are
\begin{equation}
\tilde{f}_{\text{erf}}(q) = \frac{2 \imath}{q} \exp\left( - \frac{q^2}{4 \omega^2} \right)
\end{equation}
and
\begin{equation}
\tilde{f}_{\text{erfgau}}(q) = \frac{2 \imath}{q} \exp\left( - \frac{q^2}{4 \omega^2} \right) - \frac{3^{3/2} \imath}{\omega^2} q \exp\left(- 3 \frac{q^2}{4 \omega^2} \right)
\end{equation}

\begin{figure}[h!]
\includegraphics[width=0.8\textwidth]{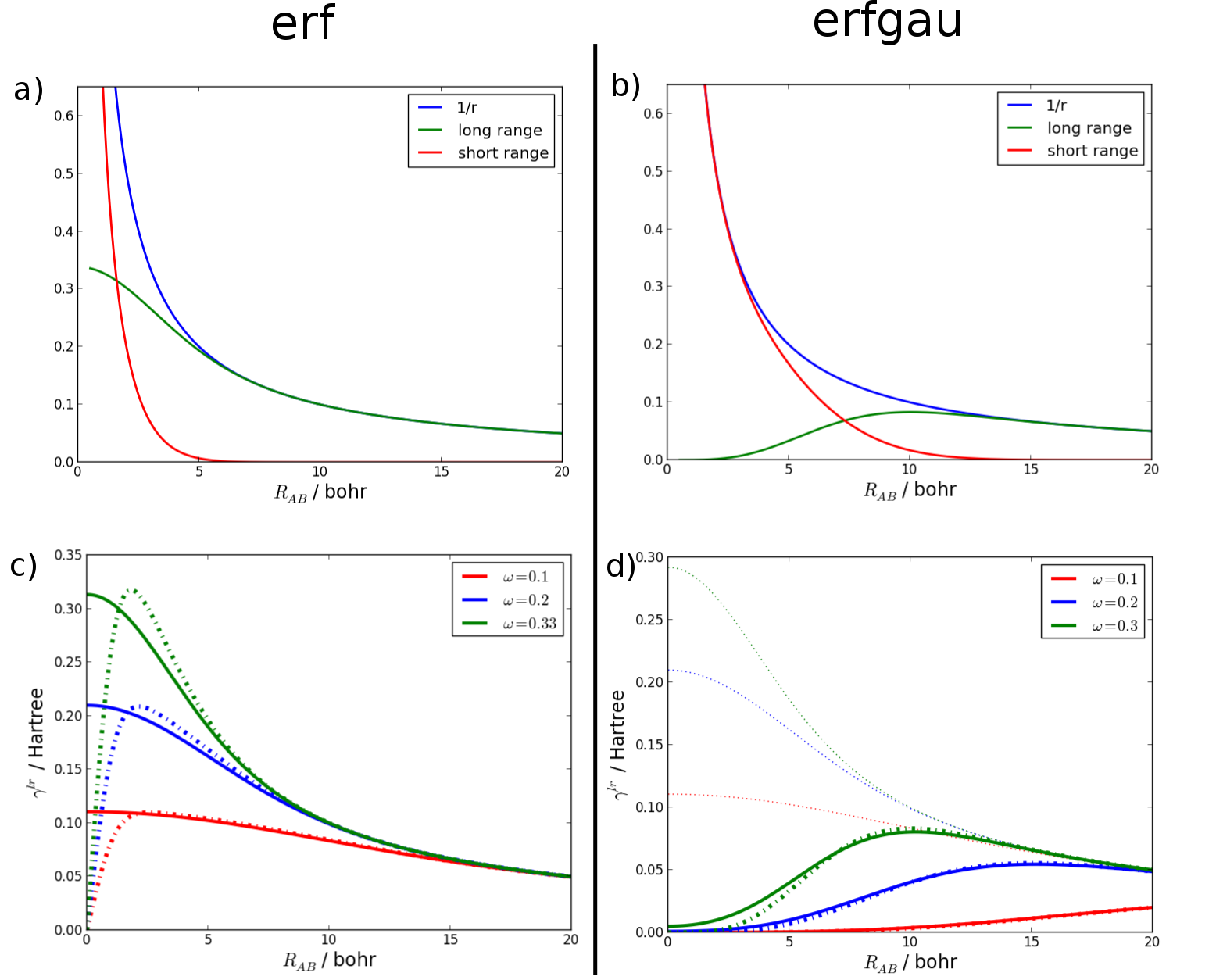}
\caption{In \textbf{a)} and \textbf{b)} the separation of the Coulomb potential into a short and long range part is depicted for the switching functions \textsl{erf} and \textsl{erfgau}, respectively. \textbf{c)} and \textbf{d)} show the resulting distance dependence of $\gamma^{lr}$ for the two switching functions. The dashed lines show the approximation according to eqn. \ref{eqn:gamma_lr_approx} and the solid lines show the numerically exact integrals \ref{eqn:gamma_lr_exact}. The thin lines in \textbf{d)} show the \textsl{erf}-integrals for comparison.}
\label{fig:gamma_lr_integrals}
\end{figure}

In Fig. \ref{fig:gamma_lr_integrals} the switching functions and $\gamma^{lr}$-integrals are compared. The short distance behaviour is completely different for \textsl{erf} and \textsl{erfgau}. 

\section{Precalculated dipole matrix elements}
\label{sec:dipoles_technical_details} 
The dipole matrix element between
two atomic orbitals $\mu$ and $\nu$ can be rewritten, so that it
depends on the relative position $\vec{R}_{\mu\nu}=\vec{R}_{\nu}-\vec{R}_{\mu}$
of the two centers and the overlap of the two orbitals: 
\begin{equation}
\begin{split}\bra{\mu(\vec{r}-\vec{R}_{\mu})} & \vec{r}\ket{\nu(\vec{r}-\vec{R}_{\nu})}=\int d^{3}r\phi_{\mu}^{*}(\vec{r}-\vec{R}_{\mu})\vec{r}\phi_{\nu}(\vec{r}-\vec{R}_{\nu})\\
 & \stackrel{\vec{r}'=\vec{r}-\vec{R}_{\mu}}{=}\int d^{3}r'\phi_{\mu}^{*}(\vec{r}')\left(\vec{r}'+\vec{R}_{\mu}\right)\phi_{\nu}(\vec{r}'+\vec{R}_{\mu}-\vec{R}_{\nu})\\
 & =\int d^{3}r'\phi_{\mu}^{*}(\vec{r}')\vec{r}'\phi_{\nu}(\vec{r}'-\vec{R}_{\mu\nu})+\vec{R}_{\mu}\int dr'^{3}\phi_{\mu}^{*}(\vec{r}')\phi_{\nu}(\vec{r}'-\vec{R}_{\mu\nu}))\label{eqn:dipole_matelem}
\end{split}
\end{equation}

The second summand in the last line of eqn. \ref{eqn:dipole_matelem}
can be calculated from Slater-Koster rules for the overlap matrix
elements between orbitals $\mu$ and $\nu$ 
\begin{equation}
\vec{R}_{\mu}\int d^{3}r'\phi_{\mu}^{*}(\vec{r}')\phi_{\nu}(\vec{r}'-\vec{R}_{\mu\nu})=\vec{R}_{\mu}S_{\mu\nu}(\vec{R}_{\mu\nu}),
\end{equation}
whereas the first summand needs special treatment as it contains the
dipole operator. The Cartesian components of the dipole operator can
be written in terms of $p_{x}$, $p_{y}$ and $p_{z}$ orbitals located
at the origin: 
\begin{eqnarray}
x & = & \sqrt{\frac{4\pi}{3}}rp_{x}\label{eqn:dipolex}\\
y & = & \sqrt{\frac{4\pi}{3}}rp_{y}\label{eqn:dipoley}\\
z & = & \sqrt{\frac{4\pi}{3}}rp_{z}\label{eqn:dipolez}
\end{eqnarray}
Therefore the dipole matrix element can be understood as the overlap
of three orbitals at two centers: 
\begin{itemize}
\item the orbital $\mu$ at the origin, 
\item a vector of p-orbitals representing the direction of the position
operator, also centered at the origin 
\item and the orbital $\nu$ at the center $\vec{R}_{\mu\nu}=\vec{R}_{\nu}-\vec{R}_{\mu}$. 
\end{itemize}
The atomic orbitals consist of a radial part $R_{nl}(r)$ and an angular
part $\tilde{Y}_{l,m}(\theta,\phi)$: 
\begin{equation}
\phi_{\mu}(\vec{r})=R_{n_{\mu}l_{\mu}}(r)\tilde{Y}_{l_{\mu},m_{\mu}}(\theta,\phi)
\end{equation}
The angular parts are real spherical harmonics: 

\begin{center}
$\begin{array}{c|c|c|c}
\text{Orbital} & l & m & \tilde{Y}_{lm}(\theta,\phi)\\
\hline s & 0 & 0 & \frac{1}{2\sqrt{\pi}}\\
p_{y} & 1 & -1 & \frac{1}{2}\sqrt{\frac{3}{\pi}}\sin(\theta)\sin(\phi)\\
p_{z} & 1 & 0 & \frac{1}{2}\sqrt{\frac{3}{\pi}}\cos(\theta)\\
p_{x} & 1 & 1 & \frac{1}{2}\sqrt{\frac{3}{\pi}}\sin(\theta)\cos(\phi)
\end{array}$ 
\par\end{center}

The radial part $R_{nl}(r)$ is specific to each atom type and is
obtained by numerically solving the radial Schrödinger equation for
the atomic Kohn-Sham Hamiltonian with a local exchange correlation
functional. For the valence $s$- and $p_{x}$,$p_{y}$ and $p_{z}$
orbitals $(l,m)$ would take the values $(0,0)$, $(1,1)$, $(1,-1)$
and $(1,0)$ respectively. The valence shell of carbon, for instance,
requires two radial functions, $R_{n=2,l=0}^{\text{C}}(r)$ for the
2s orbital, and $R_{n=2,l=1}^{\text{C}}(r)$ for the three 2p orbitals.

\begin{figure}[h!]
\includegraphics[width=0.7\textwidth]{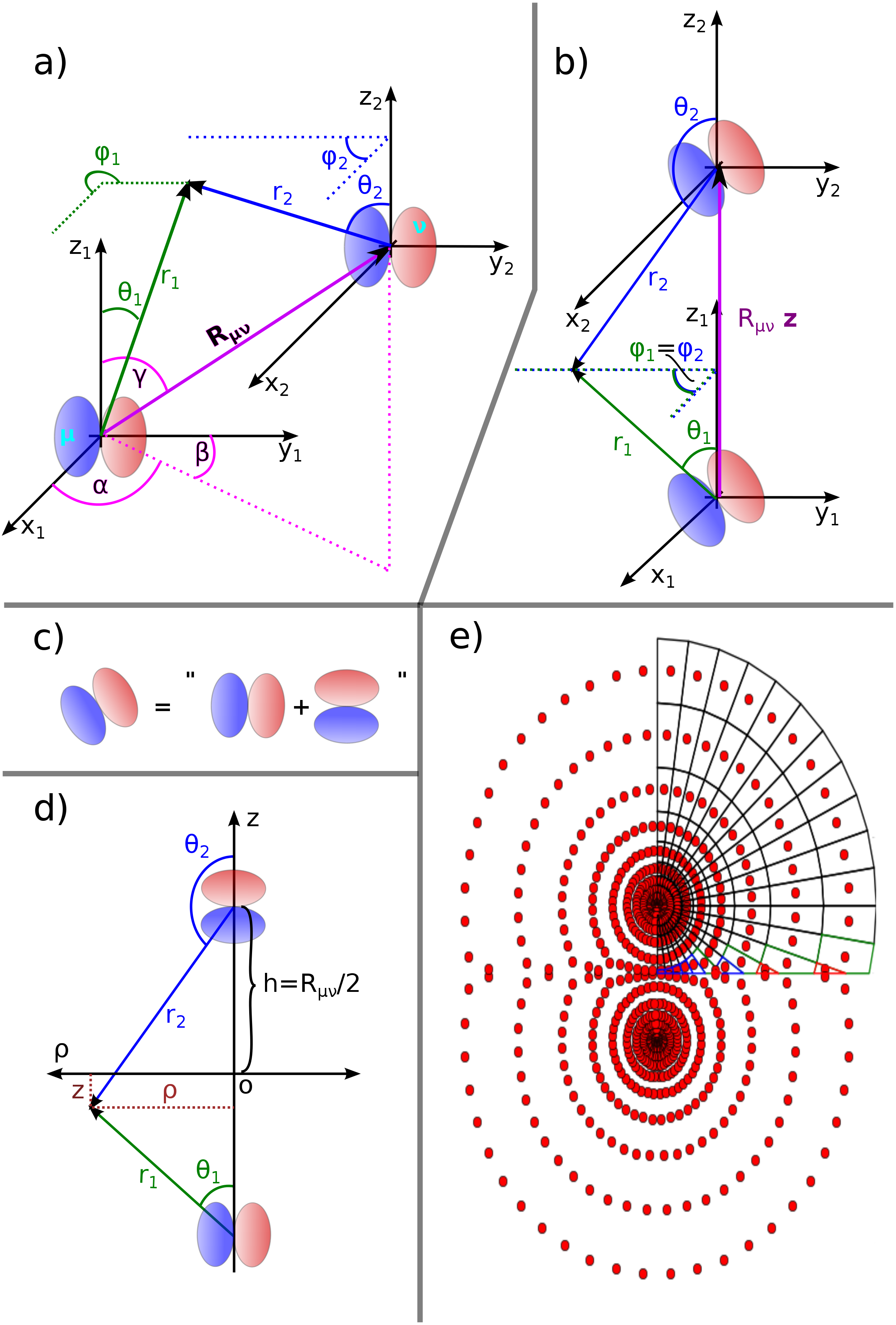} \caption{ \textbf{a)} Local coordinate systems 1 and 2 around orbitals $\mu$
and $\nu$. The spherical coordinates of a position $\vec{r}'$ can
be specified with respect to either axis. The direction of the vector
$\vec{R}_{\mu\nu}$ joining the two orbital centers is defined by
the directional cosines $x=\cos(\alpha)$, $y=\cos(\beta)$ and $z=\cos(\gamma)$.
\textbf{b)} After rotating the coordinate systems $\vec{R}_{\mu\nu}$
coincides with the z-axes. \textbf{c)} The rotated spherical harmonics
are linear combinations of spherical harmonics aligned with the axes.
\textbf{d)} Cylindrical coordinates. \textbf{e)} Grid for integration.
Two polar grids centered at the atomic positions are merged for an
efficient distribution of sampling points around both atoms (implementation
of \textsl{Hotbit}\cite{DFTB_for_beginners}). The plane is divided
into rectangles or triangles, whose size increases away from each
center. }

\label{fig:SKrotations} 
\end{figure}

Decomposing the orbitals $\mu$ and $\nu$ into their radial and angular
parts and expressing the dipole operator by eqns.\ref{eqn:dipolex},\ref{eqn:dipoley}
and \ref{eqn:dipolez}, the integral for the first summand in eqn.
\ref{eqn:dipole_matelem} becomes (in spherical coordinates): 
\begin{equation}
\begin{split}\int d^{3}r'\phi_{\mu}^{*}(\vec{r}') & \vec{r}'\phi_{\nu}(\vec{r}'-\vec{R}_{\mu\nu})=\underbrace{\int_{0}^{\infty}r_{1}^{2}dr_{1}\int_{0}^{\pi}\sin(\theta_{1})d\theta_{1}\int_{0}^{2\pi}d\phi_{1}}_{d^{3}r_{1}}\\
 & \times\underbrace{R_{n_{\mu}l_{\mu}}^{*}(r_{1})\tilde{Y}_{l_{\mu}m_{\mu}}^{*}(\theta_{1},\phi_{1})}_{\phi_{\mu}^{*}(\vec{r}_{1})}\underbrace{\sqrt{\frac{4\pi}{3}}r_{1}\begin{pmatrix}\tilde{Y}_{1,1}(\theta_{1},\phi_{1})\\
\tilde{Y}_{1,-1}(\theta_{1},\phi_{1})\\
\tilde{Y}_{1,0}(\theta_{1},\phi_{1})
\end{pmatrix}}_{\vec{r}_{1}}\underbrace{R_{n_{\nu}l_{\nu}}(\vert\vec{r}_{1}-\vec{R}_{\mu\nu}\vert)\tilde{Y}_{l_{\nu}m_{\nu}}(\theta_{2},\phi_{2})}_{\phi_{\nu}(\vec{r}_{2})}\label{eqn:integral_unrotated}
\end{split}
\end{equation}
where 
$r_{2}=\vert\vec{r}_{1}-\vec{R}_{\mu\nu}\vert$, $\theta_{2}$ and
$\phi_{2}$ depend on the integration variables $r_{1}$, $\theta_{1}$
and $\phi_{1}$ as illustrated in Fig.\ref{fig:SKrotations}a).

We wish to find a way to precalculate these integrals and tabulate
them, so that at runtime no integrals have to be solved. At first
it seems, as if one had to solve the integral for all possible relative
arrangement in 3D space (expressed by $\vec{R}_{\mu\nu}$) of the
two orbitals. Slater-Koster rules\cite{slater_koster} allow to break
the integral down to a set of few elementary integrals, that only
depend on the relative distance, and from which the integrals for
any relative orientation can be assembled quickly.

To derive the Slater-Koster rules for dipole matrix elements, we begin
by rotating the coordinate system such that $\vec{R}_{\mu\nu}$ points
along the $z$ axis. Since spherical harmonics form a representation
of the rotation group $SO(3)$ for each angular momentum $l$, the
action of this rotation is to mix spherical harmonics with different
$m$ but the same $l$. For (complex) spherical harmonics the mixing
is described by the Wigner D-matrices, analogously real spherical
harmonics, as they are used for orbitals, will be transformed by combinations
of those $D$-matrices, which will be called $\tilde{D}$-matrices:
\begin{equation}
\mathcal{R}_{\vec{R}_{\mu\nu}\to\hat{z}}\left[\tilde{Y}_{lm}(\theta,\phi)\right]=\sum_{m'=-l}^{l}\tilde{D}_{m,m'}^{l}(A,B,\Gamma)\tilde{Y}_{l,m'}(\theta,\phi)
\end{equation}
The $\tilde{D}$ matrices are expressed as functions of three Euler
angles $A,B$ and $\Gamma$. In order to align the vector $\vec{R}_{\mu\nu}$
(whose spherical coordinates are $R,\Theta$ and $\Phi$) with the
z-axis, the angles have to be set to $A=\frac{\pi}{2}$, $B=\Theta$
and $\Gamma=\Phi$. In the integral \ref{eqn:integral_unrotated}
all three spherical harmonics have to be rotated leading to: 
\begin{equation}
\begin{split}\int d^{3}r'\phi_{\mu}^{*}(\vec{r}') & \vec{r}'\phi_{\nu}(\vec{r}'-\vec{R}_{\mu\nu})=\sum_{m_{1}=-l_{\mu}}^{l_{\mu}}\sum_{m_{2}=-1}^{1}\sum_{m_{3}=-l_{\nu}}^{l_{\nu}}\underbrace{\left(\tilde{D}_{m_{\mu},m_{1}}^{l_{\mu}}\right)^{*}\begin{pmatrix}\tilde{D}_{1,m_{2}}^{1}\\
\tilde{D}_{-1,m_{2}}^{1}\\
\tilde{D}_{0,m_{2}}^{0}
\end{pmatrix}\tilde{D}_{m_{\nu},m_{3}}^{l_{\nu}}}_{T_{i(l_{\mu},l_{\nu},m_{\mu},m_{\nu},m_{1},m_{2},m_{3})}}\\
 & \times\int_{0}^{\infty}r_{1}^{2}dr_{1}\int_{0}^{\pi}\sin(\theta_{1})d\theta_{1}r_{1}R_{n_{\mu},l_{\mu}}^{*}(r_{1})R_{n_{\nu},l_{\nu}}(r_{2})\\
 & \times\underbrace{\sqrt{\frac{4\pi}{3}}\int_{0}^{2\pi}d\phi\tilde{Y}_{l_{\mu},m_{1}}^{*}(\theta_{1},\phi)\tilde{Y}_{1,m_{2}}(\theta_{1},\phi)\tilde{Y}_{l_{\nu},m_{3}}(\theta_{2},\phi)}_{\phi_{i(l_{\mu},l_{\nu},m_{1},m_{2},m_{3})}(\theta_{1},\theta_{2})}
\end{split}
\end{equation}
The rotated coordinate systems and mixing of spherical harmonics is
depicted in Fig.\ref{fig:SKrotations}b) and c).

Since after the rotation the z-axes for both orbital centered coordinate
systems coincide, one has $\phi_{1}=\phi_{2}$ and the $\phi$ integral
can be done analytically\cite{angular_momentum_in_QM}. For $s$ and
$p$ orbitals 8 different expressions $\phi_{i}(\theta_{1},\theta_{2})$
result, which are listed in the appendix \ref{sec:unique_radial_integrals}
in table \ref{tbl:phi_integrals}. The remaining two-dimensional integrals
over $r_{1}$ and $\theta_{1}$ are best performed in cylindrical
coordinates $\rho,z$. The variable transformations are given by 
\begin{eqnarray}
r_{1} & = & \sqrt{\rho^{2}+(z-h)^{2}} \label{eqn:definition_r1}\\
r_{2} & = & \sqrt{\rho^{2}+(z+h)^{2}}\\
\theta_{1} & = & \arctan2(\rho,z-h)\\
\theta_{2} & = & \arctan2(\rho,z+h) \label{eqn:definition_theta_2}
\end{eqnarray}
where $h=\frac{R_{\mu\nu}}{2}$ is half the distance between the two
atomic centers (see Fig.\ref{fig:SKrotations}d) ). Since the orbitals
are highly peaked at the atomic position and decay exponentially towards
larger distances, a grid with sampling points clustered around the
two atomic centers (see Fig. \ref{fig:SKrotations}e) ) is most suited
for accurate quadrature.

In total, one has to calculate the 8 two-dimensional integrals $D_{i}(R_{\mu\nu})$
listed in the third column of table \ref{tbl:phi_integrals} as a
function of the orbital separation and save them to a file. The coefficients
$T_{i}(x,y,z)$ depend on the directional cosines of the vector $\vec{R}_{\mu\nu}$,
i.e. $x=\cos(\alpha)$,$y=\cos(\beta)$ and $z=\cos(\gamma)$ as labelled
in Fig.\ref{fig:SKrotations}a), and account for the relative orientation
of the orbitals.

\begin{equation}
\int d^{3}r'\phi_{\mu}^{*}(\vec{r}')\vec{r}'\phi_{\nu}(\vec{r}'-\vec{R}_{\mu\nu})=\sum_{i}T_{i}(x,y,z)D_{i}(R_{\mu\nu})
\end{equation}

\begin{figure}[h!]
\includegraphics[width=0.8\textwidth]{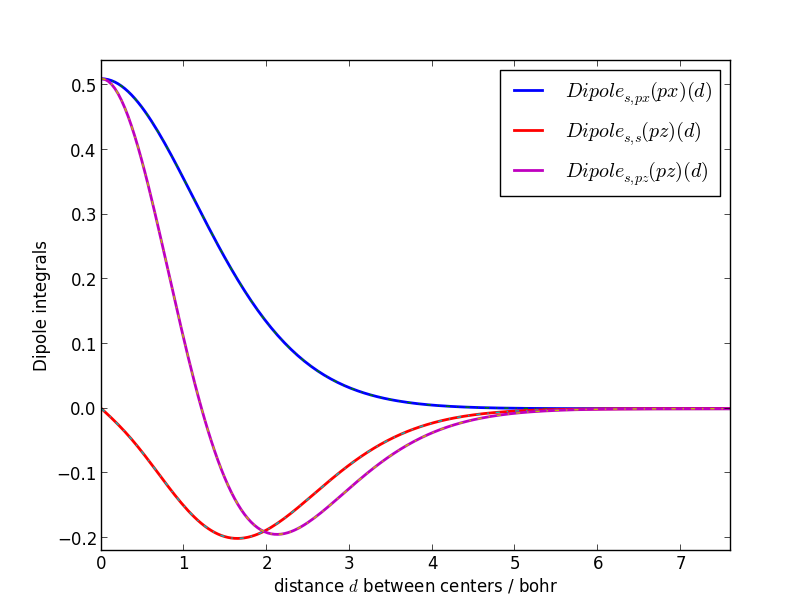} \caption{\textbf{Precalculated dipole integrals} for the atom pair h-c that
are tabulated in Slater-Koster files, $D_{1}$ (blue), $D_{2}$ (red)
and $D_{3}$ (violet). }

\label{fig:slako_dipole_h_c} 
\end{figure}

The Slater-Koster rules for computing $\sum_{i}T_{i}D_{i}$ are given
in table \ref{tbl:sk_rules} of the appendix \ref{sec:slako_rules_dipoles}.
The 3 integrals needed for dipole matrix elements between the valence
orbitals of carbon and hydrogen are shown in Fig. \ref{fig:slako_dipole_h_c}
as a function of the interatomic distance.

\FloatBarrier

\section{Unique radial integrals $D_{i}$}
\label{sec:unique_radial_integrals}

\begin{table}[h!]
\begin{tabular}{c|c|c}
$i$  & $\phi_{i}(\theta_{1},\theta_{2})$  & Integrals $D_{i}(R_{\mu\nu})$ \tabularnewline
\hline 
$1$  & $\frac{3}{8\sqrt{\pi}}\sin(\theta_{1})\sin(\theta_{2})$  & $\int\int dz\rho d\rho R_{n_{\mu},l_{\mu}={\color{blue}0}}(r_{1})r_{1}R_{n_{\nu},l_{\nu}={\color{blue}1}}(r_{2})\phi_{1}(\theta_{1},\theta_{2})$ \tabularnewline
$2$  & $\frac{\sqrt{3}}{4\sqrt{\pi}}\cos(\theta_{1})$  & $\int\int dz\rho d\rho R_{n_{\mu},l_{\mu}={\color{blue}0}}(r_{1})r_{1}R_{n_{\nu},l_{\nu}={\color{blue}0}}(r_{2})\phi_{1}(\theta_{1},\theta_{2})$ \tabularnewline
$3$  & $\frac{3}{4\sqrt{\pi}}\cos(\theta_{1})\cos(\theta_{2})$  & $\int\int dz\rho d\rho R_{n_{\mu},l_{\mu}={\color{blue}0}}(r_{1})r_{1}R_{n_{\nu},l_{\nu}={\color{blue}1}}(r_{2})\phi_{1}(\theta_{1},\theta_{2})$\tabularnewline
$4$  & $\frac{3}{8\sqrt{\pi}}\sin^{2}(\theta_{1})$  & $\int\int dz\rho d\rho R_{n_{\mu},l_{\mu}={\color{blue}1}}(r_{1})r_{1}R_{n_{\nu},l_{\nu}={\color{blue}0}}(r_{2})\phi_{1}(\theta_{1},\theta_{2})$\tabularnewline
$5$  & $\frac{3}{8}\sqrt{\frac{3}{\pi}}\sin^{2}(\theta_{1})\cos(\theta_{2})$  & $\int\int dz\rho d\rho R_{n_{\mu},l_{\mu}={\color{blue}1}}(r_{1})r_{1}R_{n_{\nu},l_{\nu}={\color{blue}1}}(r_{2})\phi_{1}(\theta_{1},\theta_{2})$\tabularnewline
$6$  & $\frac{3}{8}\sqrt{\frac{3}{\pi}}\sin(\theta_{1})\cos(\theta_{1})\sin(\theta_{2})$  & $\int\int dz\rho d\rho R_{n_{\mu},l_{\mu}={\color{blue}1}}(r_{1})r_{1}R_{n_{\nu},l_{\nu}={\color{blue}1}}(r_{2})\phi_{1}(\theta_{1},\theta_{2})$\tabularnewline
$7$  & $\frac{3}{4\sqrt{\pi}}\cos^{2}(\theta_{1})$  & $\int\int dz\rho d\rho R_{n_{\mu},l_{\mu}={\color{blue}1}}(r_{1})r_{1}R_{n_{\nu},l_{\nu}={\color{blue}0}}(r_{2})\phi_{1}(\theta_{1},\theta_{2})$\tabularnewline
$8$  & $\frac{3\sqrt{3}}{4\sqrt{\pi}}\cos^{2}(\theta_{1})\cos(\theta_{2})$  & $\int\int dz\rho d\rho R_{n_{\mu},l_{\mu}={\color{blue}1}}(r_{1})r_{1}R_{n_{\nu},l_{\nu}={\color{blue}1}}(r_{2})\phi_{1}(\theta_{1},\theta_{2})$\tabularnewline
\end{tabular}\caption{List of the angular and radial integrals. $R_{n_{\mu},l_{\mu}}(r)$
denotes the radial function with angular momentum $l_{\mu}$ for the
orbital $\mu$ of one atom type. $r_{1}$,$r_{2}$ and $\theta_{1}$
and $\theta_{2}$ depend on the cylindrical integration variables
$\rho$ and $z$ as explained in eqns. \ref{eqn:definition_r1}-\ref{eqn:definition_theta_2}.}

\label{tbl:phi_integrals} 
\end{table}

\section{Slater-Koster rules for assembling dipole matrix elements}

\label{sec:slako_rules_dipoles} Table \ref{tbl:sk_rules} summarizes
the rules for calculating dipole matrix elements between two atomic-centered
orbitals from Slater-Koster tables for the integrals $D_{i}(\vec{R}_{12})$.
The atomic orbital $\phi_{1}$ is centered on an atom at position
$\vec{R}_{1}$ and the second atomic orbital $\phi_{2}$ is centered
on another atom at position $\vec{R}_{2}$. $x,y$ and $z$ are the
directional cosines for the vector $\vec{R}_{12}$ pointing from the
first to the second center. The 8 integrals $D_{i=1,\ldots,8}(r)$
are precalculated for all distances $r=\vert\vec{R}_{12}\vert$ and
tabulated. So far only $s$- and $p$-orbital are considered. Since
the rules were obtained from a computer algebra program written for
the software package \textsl{Mathematica}\cite{Mathematica}, rules
for $d$-orbitals could also be obtained easily.

\begin{table}[h!]
{\footnotesize{}
\begin{tabular}{cccl}
{\footnotesize{orb. $\phi_{1}$ }} & {\footnotesize{coord. }} & {\footnotesize{orb. $\phi_{2}$ }} & {\footnotesize{Rule for $\bra{\phi_{1}(\vec{r}-\vec{R}_{1})}\left(\vec{r}-\vec{R}_{12}\right)\ket{\phi_{2}(\vec{r}-\vec{R}_{2})}$ }}\tabularnewline
\hline 
{\footnotesize{$s$ }} & {\footnotesize{y }} & {\footnotesize{$s$ }} & {\footnotesize{$yD_{2}(r)$ }}\tabularnewline
{\footnotesize{$s$ }} & {\footnotesize{y }} & {\footnotesize{$p_{y}$ }} & {\footnotesize{$(x^{2}+z^{2})D_{1}(r)+y^{2}D_{3}(r)$ }}\tabularnewline
{\footnotesize{$s$ }} & {\footnotesize{y }} & {\footnotesize{$p_{z}$ }} & {\footnotesize{$yz(D_{3}(r)-D_{1}(r))$ }}\tabularnewline
{\footnotesize{$s$ }} & {\footnotesize{y }} & {\footnotesize{$p_{x}$ }} & {\footnotesize{$xy(D_{3}(r)-D_{1}(r))$ }}\tabularnewline
{\footnotesize{$s$ }} & {\footnotesize{z }} & {\footnotesize{$s$ }} & {\footnotesize{$zD_{2}(r)$ }}\tabularnewline
{\footnotesize{$s$ }} & {\footnotesize{z }} & {\footnotesize{$p_{y}$ }} & {\footnotesize{$yz(D_{3}(r)-D_{1}(r))$ }}\tabularnewline
{\footnotesize{$s$ }} & {\footnotesize{z }} & {\footnotesize{$p_{z}$ }} & {\footnotesize{$(x^{2}+y^{2})D_{1}(r)+z^{2}D_{3}(r)$ }}\tabularnewline
{\footnotesize{$s$ }} & {\footnotesize{z }} & {\footnotesize{$p_{x}$ }} & {\footnotesize{$xz(D_{3}(r)-D_{1}(r))$ }}\tabularnewline
{\footnotesize{$s$ }} & {\footnotesize{x }} & {\footnotesize{$s$ }} & {\footnotesize{$xD_{2}(r)$ }}\tabularnewline
{\footnotesize{$s$ }} & {\footnotesize{x }} & {\footnotesize{$p_{y}$ }} & {\footnotesize{$xy(D_{3}(r)-D_{1}(r))$ }}\tabularnewline
{\footnotesize{$s$ }} & {\footnotesize{x }} & {\footnotesize{$p_{z}$ }} & {\footnotesize{$xz(D_{3}(r)-D_{1}(r))$ }}\tabularnewline
{\footnotesize{$s$ }} & {\footnotesize{x }} & {\footnotesize{$p_{x}$ }} & {\footnotesize{$(y^{2}+z^{2})D_{1}(r)+x^{2}D_{3}(r)$ }}\tabularnewline
\hline 
{\footnotesize{$p_{y}$}} & {\footnotesize{y }} & {\footnotesize{$s$ }} & {\footnotesize{$(x^{2}+y^{2})D_{4}(r)+y^{2}D_{7}(r)$ }}\tabularnewline
{\footnotesize{$p_{y}$}} & {\footnotesize{y }} & {\footnotesize{$p_{y}$ }} & {\footnotesize{$y(x^{2}+z^{2})(D_{5}(r)+2D_{6}(r))+y^{3}D_{8}(r)$ }}\tabularnewline
{\footnotesize{$p_{y}$}} & {\footnotesize{y }} & {\footnotesize{$p_{z}$ }} & {\footnotesize{$z((x^{2}+z^{2})D_{5}(r)+y^{2}(D_{8}(r)-2D_{6}(r)))$ }}\tabularnewline
{\footnotesize{$p_{y}$}} & {\footnotesize{y }} & {\footnotesize{$p_{x}$ }} & {\footnotesize{$x((x^{2}+z^{2})D_{5}(r)+y^{2}(D_{8}(r)-2D_{6}(r)))$ }}\tabularnewline
{\footnotesize{$p_{y}$}} & {\footnotesize{z }} & {\footnotesize{$s$ }} & {\footnotesize{$yz(D_{7}(r)-D_{4}(r))$ }}\tabularnewline
{\footnotesize{$p_{y}$}} & {\footnotesize{z }} & {\footnotesize{$p_{y}$ }} & {\footnotesize{$z((x^{2}-y^{2}+z^{2})D_{6}(r)+y^{2}(D_{8}(r)-D_{5}(r)))$ }}\tabularnewline
{\footnotesize{$p_{y}$}} & {\footnotesize{z }} & {\footnotesize{$p_{z}$ }} & {\footnotesize{$y((x^{2}+y^{2})D_{6}(r)-z^{2}(D_{5}(r)+D_{6}(r)-D_{8}(r)))$ }}\tabularnewline
{\footnotesize{$p_{y}$}} & {\footnotesize{z }} & {\footnotesize{$p_{x}$ }} & {\footnotesize{$xyz(D_{8}(r)-D_{5}(r)-2D_{6}(r))$ }}\tabularnewline
{\footnotesize{$p_{y}$}} & {\footnotesize{x }} & {\footnotesize{$s$ }} & {\footnotesize{$xy(D_{7}(r)-D_{4}(r))$ }}\tabularnewline
{\footnotesize{$p_{y}$}} & {\footnotesize{x }} & {\footnotesize{$p_{y}$ }} & {\footnotesize{$x((x^{2}-y^{2}+z^{2})D_{6}(r)+y^{2}D_{8}(r)-y^{2}D_{5}(r))$ }}\tabularnewline
{\footnotesize{$p_{y}$}} & {\footnotesize{x }} & {\footnotesize{$p_{z}$ }} & {\footnotesize{$xyz(D_{8}(r)-D_{5}(r)-2D_{6}(r))$ }}\tabularnewline
{\footnotesize{$p_{y}$}} & {\footnotesize{x }} & {\footnotesize{$p_{x}$ }} & {\footnotesize{$y(x^{2}(D_{8}(r)-D_{5}(r))+(z^{2}-x^{2}+y^{2})D_{6}(r))$ }}\tabularnewline
\hline 
{\footnotesize{$p_{z}$}} & {\footnotesize{y }} & {\footnotesize{$s$ }} & {\footnotesize{$yz(D_{7}(r)-D_{4}(r))$ }}\tabularnewline
{\footnotesize{$p_{z}$}} & {\footnotesize{y }} & {\footnotesize{$p_{y}$ }} & {\footnotesize{$z(y^{2}(D_{8}(r)-D_{5}(r))+(x^{2}-y^{2}+z^{2})D_{6}(r))$ }}\tabularnewline
{\footnotesize{$p_{z}$}} & {\footnotesize{y }} & {\footnotesize{$p_{z}$ }} & {\footnotesize{$y((x^{2}+y^{2})D_{6}(r)-z^{2}(D_{5}(r)+D_{6}(r)-D_{8}(r)))$ }}\tabularnewline
{\footnotesize{$p_{z}$}} & {\footnotesize{y }} & {\footnotesize{$p_{x}$ }} & {\footnotesize{$xyz(D_{8}(r)-D_{5}(r)-2D_{6}(r))$ }}\tabularnewline
{\footnotesize{$p_{z}$}} & {\footnotesize{z }} & {\footnotesize{$s$ }} & {\footnotesize{$(x^{2}+y^{2})D_{4}(r)+z^{2}D_{7}(r)$ }}\tabularnewline
{\footnotesize{$p_{z}$}} & {\footnotesize{z }} & {\footnotesize{$p_{y}$ }} & {\footnotesize{$y((x^{2}+y^{2})D_{5}(r)+z^{2}(D_{8}(r)-2D_{6}(r)))$ }}\tabularnewline
{\footnotesize{$p_{z}$}} & {\footnotesize{z }} & {\footnotesize{$p_{z}$ }} & {\footnotesize{$(x^{2}+y^{2})z(D_{5}(r)+2D_{6}(r))+z^{3}D_{8}(r)$ }}\tabularnewline
{\footnotesize{$p_{z}$}} & {\footnotesize{z }} & {\footnotesize{$p_{x}$ }} & {\footnotesize{$x((x^{2}+y^{2})D_{5}(r)+z^{2}(D_{8}(r)-2D_{6}(r)))$ }}\tabularnewline
{\footnotesize{$p_{z}$}} & {\footnotesize{x }} & {\footnotesize{$s$ }} & {\footnotesize{$xz(D_{7}(r)-D_{4}(r))$ }}\tabularnewline
{\footnotesize{$p_{z}$}} & {\footnotesize{x }} & {\footnotesize{$p_{y}$ }} & {\footnotesize{$xyz(D_{8}(r)-D_{5}(r)-2D_{6}(r))$ }}\tabularnewline
{\footnotesize{$p_{z}$}} & {\footnotesize{x }} & {\footnotesize{$p_{z}$ }} & {\footnotesize{$x((x^{2}+y^{2})D_{6}(r)-z^{2}(D_{5}(r)+D_{6}(r)-D_{8}(r)))$ }}\tabularnewline
{\footnotesize{$p_{z}$}} & {\footnotesize{x }} & {\footnotesize{$p_{x}$ }} & {\footnotesize{$z(x^{2}(D_{8}(r)-D_{5}(r))+(y^{2}-x^{2}+z^{2})D_{6}(r))$ }}\tabularnewline
\hline 
{\footnotesize{$p_{x}$}} & {\footnotesize{y }} & {\footnotesize{$s$ }} & {\footnotesize{$xy(D_{7}(r)-D4(r))$ }}\tabularnewline
{\footnotesize{$p_{x}$}} & {\footnotesize{y }} & {\footnotesize{$p_{y}$ }} & {\footnotesize{$x(y^{2}(D_{8}(r)-D_{5}(r))+(x^{2}-y^{2}+z^{2})D_{6}(r))$ }}\tabularnewline
{\footnotesize{$p_{x}$}} & {\footnotesize{y }} & {\footnotesize{$p_{z}$ }} & {\footnotesize{$xyz(D_{8}(r)-D_{5}(r)-2D_{6}(r))$ }}\tabularnewline
{\footnotesize{$p_{x}$}} & {\footnotesize{y }} & {\footnotesize{$p_{x}$ }} & {\footnotesize{$y(x^{2}(D_{8}(r)-D_{5}(r))+(y^{2}-x^{2}+z^{2})D_{6}(r))$ }}\tabularnewline
{\footnotesize{$p_{x}$}} & {\footnotesize{z }} & {\footnotesize{$s$ }} & {\footnotesize{$xz(D_{7}(r)-D_{4}(r))$ }}\tabularnewline
{\footnotesize{$p_{x}$}} & {\footnotesize{z }} & {\footnotesize{$p_{y}$ }} & {\footnotesize{$xyz(D_{8}(r)-D_{5}(r)-2D_{6}(r))$ }}\tabularnewline
{\footnotesize{$p_{x}$}} & {\footnotesize{z }} & {\footnotesize{$p_{z}$ }} & {\footnotesize{$x((x^{2}+y^{2})D_{6}(r)-z^{2}(D_{5}(r)+D_{6}(r)-D_{8}(r)))$ }}\tabularnewline
{\footnotesize{$p_{x}$}} & {\footnotesize{z }} & {\footnotesize{$p_{x}$ }} & {\footnotesize{$z(x^{2}(D_{8}(r)-D_{5}(r))+(y^{2}-x^{2}+z^{2})D_{6}(r))$ }}\tabularnewline
{\footnotesize{$p_{x}$}} & {\footnotesize{x }} & {\footnotesize{$s$ }} & {\footnotesize{$(y^{2}+z^{2})D_{4}(r)+x^{2}D_{7}(r)$ }}\tabularnewline
{\footnotesize{$p_{x}$}} & {\footnotesize{x }} & {\footnotesize{$p_{y}$ }} & {\footnotesize{$y((y^{2}+z^{2})D_{5}(r)+x^{2}(D_{8}(r)-2D_{6}(r)))$ }}\tabularnewline
{\footnotesize{$p_{x}$}} & {\footnotesize{x }} & {\footnotesize{$p_{z}$ }} & {\footnotesize{$z((y^{2}+z^{2})D_{5}(r)+x^{2}(D_{8}(r)-2D_{6}(r)))$ }}\tabularnewline
{\footnotesize{$p_{x}$}} & {\footnotesize{x }} & {\footnotesize{$p_{x}$ }} & {\footnotesize{$x(y^{2}+z^{2})(D_{5}(r)+2D_{6}(r))+x^{3}D_{8}(r)$ }}\tabularnewline
\hline 
\end{tabular} 
}
\caption{Slater-Koster rules for dipole matrix elements (for s and p-orbitals).
For clarity the distance between orbital centers is named $r$ instead
of $R_{\mu\nu}$.}

\label{tbl:sk_rules} 
\end{table}

\FloatBarrier

\section{Analysing charge transfer with density differences}

\label{sec:ct_analysis_drho} 
% CT metric
Different ``metrics'' have been devised to identify charge-transfer states automatically and warn about possible failures of TD-DFT. 
Since the $\Lambda$-metric\cite{lambda_diagnostic} cannot detect all problematic charge transfer excitations, Guido et.al. introduced the $\Delta r$\cite{Guido_dr_metric}- and $\Gamma$\cite{Guido_Gamma_metric}-metrics. In particular the $\Delta r$-metric has an intuitive interpretation as the electron-hole distance. 

We can define a similar quantity at the tight-binding level. 
To this end, we start with the density difference between the excited
state and the ground state, $\Delta\rho_{I}=\rho_{I}-\rho_{0}$. 
In the linear response regime the Kohn-Sham ``wavefunction'' of the
excited state $I$ is a linear combination of single excitations from
the Kohn-Sham ground state Slater determinant: 
\begin{equation}
\ket{\Psi_{I}}=\sum_{v\in\text{virt}}\sum_{o\in\text{occ}}C_{vo}^{I}\hat{a}_{v}^{\dagger}\hat{a}_{o}\ket{\Psi_{0}}\label{eqn:CIS}
\end{equation}

The operator for the electron density in second quantization reads
(in the basis of Kohn-Sham orbitals): 
\begin{equation}
\hat{\rho}(\vec{r})=\sum_{\alpha}\sum_{\beta}\hat{a}_{\alpha}^{\dagger}\hat{a}_{\beta}\phi_{\alpha}^{*}(\vec{r})\phi_{\beta}(\vec{r})\label{eqn:rho_2ndQ}
\end{equation}

Here, $o$ and $o'$ denote occupied, $v$ and $v'$ virtual and $\alpha$
and $\beta$ general Kohn-Sham orbitals. Combining eqns. \ref{eqn:CIS}
and \ref{eqn:rho_2ndQ}, the density of state $I$ becomes: 
\begin{equation}
\rho_{I}(\vec{r})=\bra{\Psi_{I}}\hat{\rho}\ket{\Psi_{I}}=\sum_{o,o'}\sum_{v,v'}\sum_{\alpha,\beta}\phi_{\alpha}^{*}(\vec{r})\phi_{\beta}(\vec{r})C_{vo}^{I*}C_{v'o'}^{I}\bra{\Psi_{0}}\hat{a}_{o}^{\dagger}\hat{a}_{v}\hat{a}_{\alpha}^{\dagger}\hat{a}_{\beta}\hat{a}_{v'}^{\dagger}\hat{a}_{o'}\ket{\Psi_{0}}
\end{equation}
By using anti-commutation relations for Fermions, and the fact that
the ground state only contains occupied orbitals (so $\hat{a}_{v}\ket{\Psi_{0}}=\hat{a}_{o}^{\dagger}\ket{\Psi_{0}}=0$),
the expression for the density can be reduced to \cite{exciton_diffusion}:
\begin{eqnarray}
\rho_{I}(\vec{r}) & = & \sum_{o}\sum_{v,v'}C_{vo}^{I*}C_{v'o}^{I}\phi_{v}^{*}(\vec{r})\phi_{v}(\vec{r})-\sum_{v}\sum_{o,o'}C_{vo}^{I*}C_{vo'}^{I}\phi_{o'}^{*}(\vec{r})\phi_{o}(\vec{r})+\sum_{o}\vert\phi_{o}(\vec{r})\vert^{2}\label{eqn:rho_I}\\
 & = & \rho_{p}(\vec{r})+\rho_{h}(\vec{r})+\rho_{o}(\vec{r})
\end{eqnarray}
In the exciton picture, the $I$-th excited state can be described
by a bound particle-hole pair. The first term in eqn. \ref{eqn:rho_I}
can be identified with the particle, the second term belongs to the
hole, while the last term is just the ground state density. The difference
density between the ground and excited state is the sum of the particle
and hole densities: 
\begin{equation}
\Delta\rho_{I}=\rho_{I}-\rho_{0}=\rho_{e}+\rho_{h}
\end{equation}
Since tight-binding DFT deals with partial charges instead of a continuous
density distribution, we have to coarse grain the density to an atomic
resolution. Again, the transition charges are approximated as a sum
of spherically symmetric charge distributions centered on the individual
atoms: 
\begin{eqnarray}
\phi_{v}^{*}(\vec{r})\phi_{v'}(\vec{r}) & = & \sum_{A}q_{A}^{vv'}F_{A}(\vert\vec{r}-\vec{R}_{A}\vert)\\
\phi_{o}^{*}(\vec{r})\phi_{o'}(\vec{r}) & = & \sum_{A}q_{A}^{oo'}F_{A}(\vert\vec{r}-\vec{R}_{A}\vert)
\end{eqnarray}
The particle and hole densities are partitioned into atomic contributions,
\begin{eqnarray}
\rho_{e} & = & \sum_{A}\sum_{o,v,v'}C_{vo}^{I*}C_{v'o}^{I}q_{A}^{vv'}F_{A}(\vert\vec{r}-\vec{R}_{A}\vert)=\sum_{A}q_{A}^{e}F_{A}(\vert\vec{r}-\vec{R}_{A}\vert)\\
\rho_{h} & = & \sum_{A}\sum_{v,o,o'}C_{vo}^{I*}C_{vo'}^{I}q_{A}^{oo'}F_{A}(\vert\vec{r}-\vec{R}_{A}\vert)=\sum_{A}q_{A}^{h}F_{A}(\vert\vec{r}-\vec{R}_{A}\vert)
\end{eqnarray}

$q_{A}^{e}$ and $q_{A}^{h}$ are the particle-hole charges.

The average positions of the particle and the hole result from the
weighted average of all charges. 
\begin{eqnarray}
\vec{r}_{e} & = & \frac{\sum_{A}q_{A}^{e}\vec{R}_{A}}{\sum_{A}q_{A}^{e}}\\
\vec{r}_{h} & = & \frac{\sum_{A}q_{A}^{h}\vec{R}_{A}}{\sum_{A}q_{A}^{h}}
\end{eqnarray}

\begin{figure}[h!]
\includegraphics[width=0.8\textwidth]{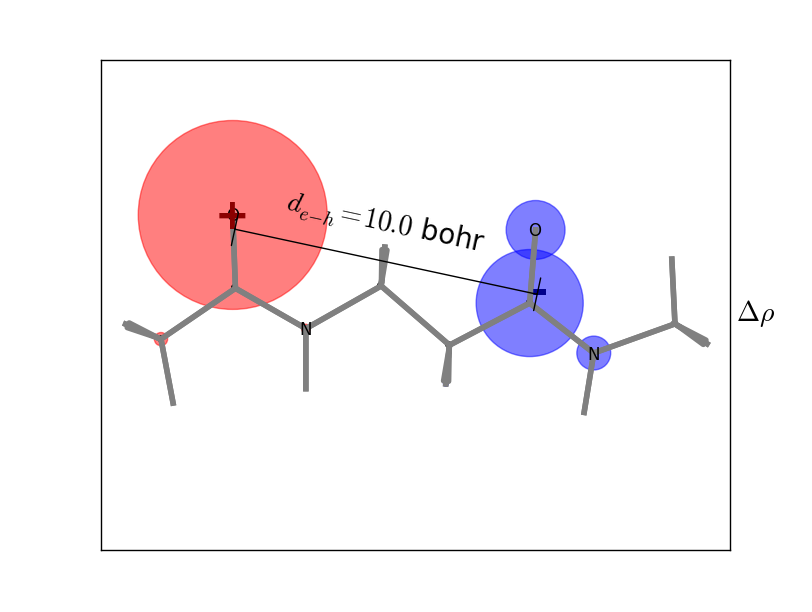}
\caption{\textbf{Difference density} $\Delta\rho$ for the 3rd excited state
of the $\beta$-dipeptide. An exciton is formed as an electron jumps
from the first carbonyl group to the second leaving a positive hole
behind. The radii of the red (blue) circles are proportional to the
hole (particle) charges on each atom.}
\label{fig:phdistance_beta-dipeptide}
\end{figure}

The particle-hole separation $d_{e-h}=\vert\vec{r}_{e}-\vec{r}_{h}\vert$
indicates the spatial extent of the exciton. Fig.\ref{fig:phdistance_beta-dipeptide}
shows that in the $\beta$-dipeptide, absorption of a photon can lead
to a separation of positive and negative charge over a distance of
10.0 bohr.

\FloatBarrier

\section{Supporting Information}
A more extensive assessment of the quality of lc-TD-DFTB excitation energies is given
in the Supporting Information. 
% lowest 10 excited singlet states of a benchmark set of 28 organic molecules
% comparison ``conventional'' TD-DFTB with PBE and the long-range corrected version with LC-PBE
This material is available free of charge via the Internet at http://pubs.acs.org
%%%% SORTED %%%%%%%%%%%%%%

\end{document}